\documentclass[10pt]{iopart}

\usepackage{bm}
\usepackage{footnote}
\usepackage{threeparttable}

\usepackage{float}

\usepackage[utf8]{inputenc}
\usepackage{cite}
\usepackage{graphicx}
\usepackage{upgreek}
\usepackage[english]{babel}
\usepackage{units}
\usepackage{booktabs}
\graphicspath{{pics/}}
\usepackage{longtable}
\usepackage{url}
\usepackage{soul}
\usepackage[breaklinks=true,colorlinks=true,linkcolor=blue,urlcolor=blue,citecolor=blue]{hyperref}
\usepackage{ntheorem}
\usepackage{multirow}
\usepackage[textwidth=3cm,textsize=scriptsize,shadow]{todonotes}

\usepackage{color} % for remark only

\newcommand*{\ja}{\textcolor{black}}

\begin{document}

\title[]{Nanosecond pulsed barrier discharge in argon for various frequencies and theoretical analysis of 2p states density ratios for $E/N$ determination}

\author{L. Kus\'yn$^{1,\star}$, A. P. Jovanovi\'c$^{2,\star}$,  D.~Loffhagen$^2$,\\ 
M. M. Becker$^{2,\wp}$ and T. Hoder$^{1,\wp}$}

\address{$^1$Department of Plasma Physics and Technology, Faculty of Science, Masaryk University,
	Kotl\'a\v{r}sk\'a 2, 61137 Brno, Czech Republic}
\address{$^2$Leibniz Institute for Plasma Science and Technology, Felix-Hausdorff-Str. 2, 17489 Greifswald, Germany}

\address{$^{\star}$These authors contributed equally to this work and share first authorship.}
\address{$^{\wp}$Corresponding authors.}

\ead{hoder@physics.muni.cz, markus.becker@inp-greifswald.de}
\vspace{10pt}

\begin{abstract}

Nanosecond pulsed barrier discharges in atmospheric pressure argon are simulated using spatially one- and two-dimensional fluid-Poisson models using the reaction kinetics model presented by Stankov {\it et al.} \cite{stankov2022}, which considers all ten argon 2p states (Paschen notation) separately. 
The very first (single) discharge  and repetitive discharges with frequencies from 5\,kHz to 100\,kHz are considered and
%The parameters of electric field, electron density, streamer velocity and electrical current are compared. 
%Obtained numerical data enable to analyse the development of 1s metastable and all 2p state source terms using the effective lifetime parameter. 
a semi-automated procedure is utilized to find appropriate 2p states for electric field determination using an intensity ratio method. 
%. 
%The method is based on evaluation of fast intensity ratio development of relative densities of two 2p states of atomic argon  
%using apparatus with high temporal resolution 
%and ion of intensity ratio development 
The proposed method is based on a \ja{time-dependent} collisional-radiative model 
%%%\ja{enabling a sub-nanosecond plasma diagnostics, it is }linking 
\ja{enabling a sub-nanosecond plasma diagnostics, it links} 
the 2p state density ratios to 
the reduced electric field strength 
$E/N$ by quantifying the excitation rate coefficients 
%(for all 2p states from Ar ground state) via solving the steady-state electron Boltzmann equation 
and by computing \ja{2p states' effective} lifetimes from the fluid model simulation. % and analysing their temporal development. 
%
%to quantify %the dependence of measured intensity ratio on 
%the electric field strength. 
%Local mean energy approximation fluid model of sinusoidal driven barrier discharge with kinetic scheme including all ten $2p$ states separately is used to exemplify the methodology. 
%The selection of effective quenching coefficients for investigated 2p states from theory and experiments is discussed as well. 
The semi-automated procedure identifies several candidates for determination of $E/N$ from given temporal profiles of the 2p state densities. 
Different approaches for effective lifetime determination are tested and applied also to measured data. 
%The E/N values computed from the selected 2p states using the simplified model are compared to numerical simulations and applied to experimental data as well. 
The influence of radial and axial 2p state density integration on the intensity ratio method is discussed. 
\ja{The above mentioned models and procedures result in a flexible theory-based methodology applicable for development of new diagnostic techniques. } 
%This intensity ratio method enables to reveal ultra-short local electric field variations in rapidly changing plasmas, such as nanosecond pulsed, dielectric barrier or filamentary streamer discharges.

%A time-dependent collision-radiative model based method is proposed theoretically for determination of the electric field strength in argon plasmas with sub-nanosecond resolution. The method development is based on the results of computer simulations. 

\end{abstract}

%\submitto{\PSST}
%\maketitle

%\pacs{52.70.-m, 52.40.Hf, 52.80.-s }% PACS, the Physics and Astronomy
%\noindent{\it Keywords}: argon, barrier discharge, reaction kinetics, time-dependent collision-radiative model, nanosecond pulsed discharge, electric field, optical emission spectroscopy, \ja{sub-nanosecond diagnostics}

%\submitto \PSST

\maketitle

\ioptwocol

\section{Introduction}
%\label{}

The electric field 
is a fundamental parameter for gas discharges and generated transient plasmas. 
The detailed knowledge of this physical quantity is important, \ja{for detailed} understanding of studied plasmas and for validation of theoretical models focused on the plasma dynamics or chemical kinetics. 
The determination of the electric field parameter is therefore a crucial task and a variety of methods is available \cite{goldberg2022} also for argon plasmas, which are of high interest in various applications \cite{massin2012,adamovich2017,desjardins2018,reuter2018,loffhagen2020}. For example, laser-induced fluorescence-dip spectroscopy \cite{czar1998} was used to determine the electric field development in lower pressure argon \cite{barnat2004}. The widely applied 
%%second harmonics based laser spectroscopy (EFISH) 
\ja{electric-field-induced second harmonics (EFISH) based laser spectroscopy} 
was recently used for investigation of plasma jets \cite{goldberg2019}, too. 
Optical emission spectroscopy (OES) based  Stark shift measurements enabled the experimental determination of the electron temperature, which is closely related to the electric field, in an atmospheric pressure argon micro-discharge in \cite{du2012}. 
A collisional-radiative model (CRM) in combination with OES measurements was used to determine the electron temperature in low pressure plasma using the line-ratio method \cite{siepa2014}. 
Line-ratio (a spectral line intensity ratio) method was recently introduced by Dyatko {\it et al.} \cite{dyatko2021} for the ionization wave electric field estimation in low pressure argon discharges. Nevertheless, the spatiotemporally highly resolved determination of 
\ja{the reduced electric field} 
$E/N$ in atmospheric pressure argon plasma using an OES method, similar to what is known for air discharges \cite{kozlov2001,bonaventura2011,hoder2016,jansky2021}, 
%(where intensity ratio of molecular nitrogen spectral bands is used), 
is not available. 
\ja{Here, $E$ is the electric field strength and $N$ denotes the number  density 
of the background gas.}

Argon barrier discharges are widely used in multiple arrangements, exhibiting broad values of parameters (electric field, electron number density, etc.) under different experimental conditions 
(\ja{applied voltage,} 
frequency, wave form, etc.). 
It is therefore desirable to have a method which is  able to access the $E/N$ under most of the assumed condition. 
For example, in the case of nanosecond pulsed applied voltage operation at different pulse frequencies. 
It is known that the discharge mechanisms can vary significantly, as it was shown e.g.\ for nitrogen-oxygen mixtures in \cite{hoder2012} or as it is apparent comparing the low frequency barrier discharge with a high frequency mode \cite{jiang2013} or even in a modulated dual-frequency mode  \cite{magnan2020,bazinette2020}. 
The intensity ratio diagnostic method of such discharges, based on \ja{simple CRM} and using an OES experiment, could be related to the emission from radiative states of the 2p manifold of argon as it usually dominates the argon barrier discharge spectra. 
As the excitation energy thresholds for 2p states differ relatively only slightly, their use for the intensity ratio method is expected to be challenging. 
Nevertheless, the experimental results presented in \cite{kusyn2023} report clearly that 2p state intensity ratios are  sensitive to the spatiotemporal dynamics of the nanosecond pulsed discharge, if recorded with high sensitivity and high temporal resolution. 

In this article, we present a  methodology how to approach the problem: by semi-automated analysis of a large amount of data obtained from numerical simulations using a time-and space-dependent fluid-Poisson modelling approach. 
With this, we investigate the possibility to use the argon atomic line intensity ratio method under above described conditions. 
\ja{The theoretical investigation provides insights into the sensitivity of the selected reaction kinetic processes for the given discharge conditions, which is also essential to quantify the limitations of the suggested 
line intensity ratio method.} 
\ja{It is worth noting that we evaluate the intensity ratio method using a time-dependent CRM, enabling possible sub-nanosecond experimental insight into argon plasmas, if fast detectors are used. The introduced and utilized methodology for intensity ratio method investigation can be applied also to other gas mixtures under various conditions, e.g. for $E/N$ determination in planetary atmospheres \cite{romero2019} or in plasmas for gas conversion \cite{heijkers2020}.}

The mentioned intensity ratio methods for electric field 
%(or mean electron energy) 
determination are designed to identify the contribution of relaxed electron ensemble given by the electron energy distribution function to two excitation processes. 
This method has previously been used for air plasmas 
%More details of the method for air plasmas can be found in 
\cite{kozlov2001,hoder2016,goldberg2022}. 
The named excitation processes populate two radiative states having different energy thresholds which are responsible for the detected optical emission. 
One of the main requirements of the method is that both excitation processes start from the same lower state, typically direct electron impact 
\ja{excitation} 
from the ground state of the atom or molecule of the 
\ja{background} 
% utilized 
gas.
%, as the density of the ground  state is known, it is almost identical to the gas density. 
This condition is fulfilled for weakly ionised low-frequency plasmas and/or under conditions with very strong collisional quenching, as it is the case in air. 
Under such conditions, nitrogen metastable states' densities decrease very fast due to the effective quenching by oxygen after each discharge and cannot contribute significantly via stepwise excitation processes as they are basically not present. 
Apparently, this condition cannot be fulfilled for repetitive or long-duration discharges in pure argon or pure nitrogen \cite{zhu2010,bilek2022,jovanovic2023,stankov2022}, where densities of metastable states become significant. To overcome this issue one needs to determine the metastable density and \ja{take the stepwise excitation process into account in the CRM related to the intensity ratio method}
\cite{mrkvickova2023}.

\ja{A second}
%Another 
% I'd prefer to repeat "a second" and to start a new paragraph to make things more clear.  
possibility is to find two radiative states whose population processes are not sensitive to the presence of the metastable states so that stepwise excitation can be neglected, at least for some conditions. 
This second possibility is theoretically more demanding yet also more experiment-friendly, as it does not require the knowledge of the metastable states' density in the studied plasma (which \ja{needs to} be additionally measured). 
In this article, we follow this second direction. 
%The result of such effort can be a robust method for plasma diagnostics.

An important part of 
%the 
intensity ratio methods for electric field determination is the knowledge of the dependence of the intensity ratio $R(E/N) $ on the reduced electric field. 
Theoretically, the ratio of two reaction rates has to be determined and the 
\ja{corresponding} 
%simplified 
simple 
\ja{CRM}
%collisional-radiative model 
has to be well founded, as it was done for air using sensitivity analysis and uncertainty quantification in \cite{obrusnik2018}. 
For plasmas in air or nitrogen, the $R(E/N)$  dependence was obtained both experimentally and theoretically e.g.\ in  \cite{goldberg2022,paris2005,obrusnik2018,bilek2018,bilek2019} and references therein. 
Paris {\it et al.} \cite{paris2005} determined the $R(E/N)$ dependence experimentally for air. 
\ja{In all these cases,}  
%%%%In all cases however, % delete however
the procedure is based on Townsend discharges with reaction kinetics in equilibrium. 
The Townsend discharge is a steady-state discharge with different properties if compared with streamer discharges, where the method is usually used. 
The exception is to additionally modify the dependence of Paris {\it et al.} \cite{paris2005} as suggested in \cite{jogi2016,hoder2016,brisset2019,bilek2018}. 
Such modified curve was used for diagnostics of streamer or nanosecond pulsed discharges \cite{jansky2021,dijcks2023} and is well within the uncertainty interval discussed in \cite{obrusnik2018,bilek2018}.
To overcome the issue of the $R(E/N)$ dependence coming from the equilibrated steady-state Townsend discharge, the method may be developed and studied also in a time-dependent case by a theoretical means (see also \cite{bonaventura2011}), i.e.\ within an argon barrier discharge simulation using an appropriate reaction kinetics model \ja{(RKM, see \cite{stankov2022} for example)} in fine spatiotemporal resolution, as we do in the present manuscript.

%About the method for ultra-fast discharges, nanosecond pulsed and streamer ones. The missing non-steady-state calibration... and the necessity of the detailed theoretical knowledge of the fast plasma processes in the investigated discharge. 
%the ground state is a good reference, the state with highest density under given conditions.  

%Here, we 
We investigate the possibility to develop an intensity ratio method for atmospheric pressure argon plasmas using numerical simulations of a barrier discharge. 
%In a recent years the team of co-authors developed an enhanced reaction kinetics model \cite{stankov2022} and obtained a first experimental results on the nanosecond pulsed barrier discharge in pure argon \cite{kusyn2023}.
Up to now, we have developed a solid theoretical foundation by creating an enhanced reaction kinetics model (see \cite{stankov2022}) used for spatially one-dimensional (1D) modelling, by developing a spatially two-dimensional (2D) fluid model for case-specific computer simulations (see \cite{jovanovic2021,jovanovic2022,jovanovic2023}) and by performing first 2p spectra measurements with sufficiently high temporal and spatial resolution to establish the link to the experiment (see \cite{kusyn2023}). 
Here, a dielectric barrier discharge in an  asymmetrical arrangement (only one electrode is covered by a dielectrics) \ja{and operated by} nanosecond pulsed voltage waveform at various frequencies is studied using numerical simulations. 
As a result, we have direct theoretical access to the development of the 
number densities of all important states and therefore to the sub-nanosecond spectra of the discharge in the most intense spectral region between 650\,nm and 900\,nm. 
%Based on the detailed model output and after careful analysis of the initiated argon plasma chemistry, we select suitable 2p states for further investigation and simplify the kinetic model to the dominant processes. 
Using a semi-automated procedure, we attempt to identify the most suitable 2p states for the intensity ratio method based on the analysis of the theoretical data and apply these findings to the experimental results.
%and propose a new intensity ratio method for determination of the electric field strength development in argon plasmas. 
%In practice, emission intensities of atomic lines resulting from the radiative depopulation of selected 2p states can be measured with such resolution and sensitivity using streak and picosecond gated cameras or using fast photomultipliers and time-correlated single photon counting techniques. 
%We also investigate the use of experimentally available effective lifetimes of the radiative states neccessary for the CRM.

The manuscript is structured in the following way. In section~\ref{sec:Modell}, the modelling procedure is described, including the \ja{RKM}~\cite{stankov2022} 
with necessary reference to our previous work. In section~\ref{Results}, the results are presented and discussed. Including the discharge dynamics under the nanosecond pulsed voltage waveform at different frequencies in subsection~\ref{sec:different_frequencies}. 
In such way, the studied discharge is well prepared for the analysis of the 2p states' kinetics in subsection~\ref{sec:semiautomatedprocedure}, where the intensity ratio method is investigated for all combinations of 2p state ratios. 
In subsection~\ref{sec:Comparison}, the effects of the density/signal integration are discussed, which are of importance for the laboratory experiment providing line-of-sight integrated profiles. 
Furthermore, a selected 2p state intensity ratio is analyzed for the measured data and compared to the results of the fluid simulations. In the final section, the presented work is concluded and summarized.

\section{Discharge arrangement and modelling procedure}\label{sec:Modell}

\begin{figure*}[!t]
\centering
%a)
%\includegraphics[clip = true,width=0.45\columnwidth]{voltages.png}
%b)
%\includegraphics[clip = true,width=0.4\columnwidth]{model01.png}
%\includegraphics[scale=1]{Figure1a.pdf}
%\includegraphics[scale=1]{Figure1b.pdf}
\includegraphics[scale=0.9]{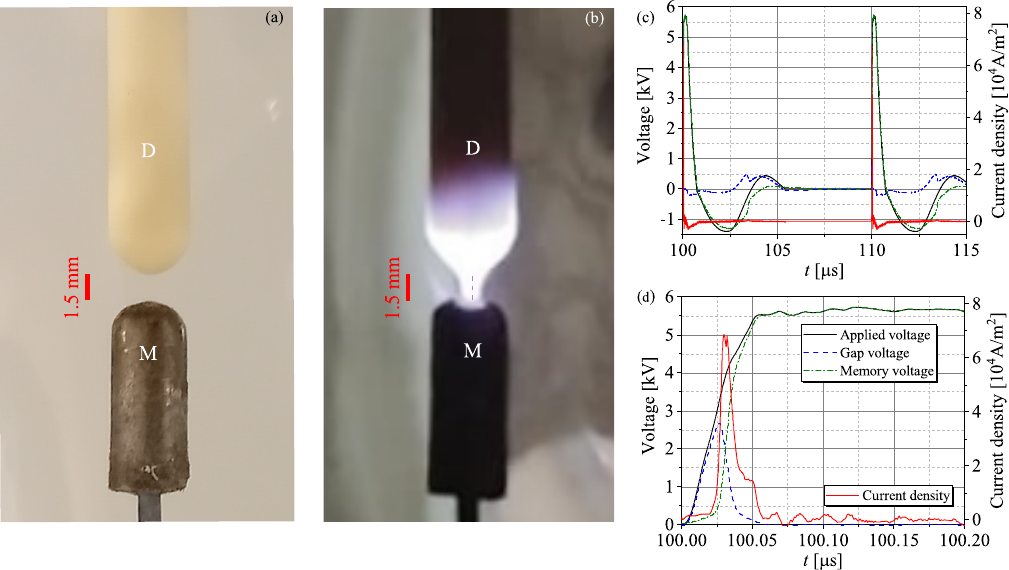}
%\end{center}
\caption{
Illustration of the asymmetric barrier discharge in atmospheric pressure argon showing (a) the photograph of the discharge geometry \ja{without discharge, (b) with discharge, and (c) and (d)} the temporal evolution of the current and voltage  after reaching periodic regime resulting from the 1D 
\ja{simulation}. 
%model. 
The metal electrode and the dielectric surface are denoted as M and D, respectively, \ja{in (a) and (b)}. %, where ``$+$â?? sign indicates the temporary anode and ``$-$â?? the temporary cathode. 
The grey dashed line in (b) illustrates the axial cut at which the 1D model was used. The part \ja{(d) zooms into the electrical parameters of the discharge event} at the rising slope of the applied voltage. \ja{The rise time (10 to 90\%) of the voltage pulse is approximately 40\,ns.}
%\DLR{Remark by DL: rise time of applied voltage and frequency are missing.}
}
\label{geometry}
\end{figure*}

Figure~\ref{geometry} shows the discharge arrangement together with typical voltage and current waveforms in the repetitive regime. 
The electrode arrangement considered in the \ja{simulations} was the same \ja{as in previous experimental and theoretical investigations \cite{hoder2011, kusyn2023, jovanovic2023}}. An asymmetric 
%\ja{To make it consistent, I would suggest to remove "dielectric" here, or, alternatively, to add it everywhere else.} 
\ja{volume} barrier discharge with one electrode covered by the alumina dielectric (96\% purity Al$_2$O$_3$) is studied. 
\ja{The electrodes are made of stainless steel.}
The gas gap is 1.5\,mm, 
the dielectric thickness is 
\ja{0.5\,mm (in \cite{hoder2011}, wrong value of the dielectric thickness of 1\,mm) is given for the same setup}, 
%\DLR{Remark by DL: dielectric thickness is 1.0\,mm in \cite{hoder2011}, 
%no value is given in \cite{kusyn2023} and 0.5\,mm was used in \cite{jovanovic2023} as well as in \cite{becker2013}!}
the dielectric's permittivity is $\varepsilon_r$ = 9, and the radius of the electrode surfaces is 2\,mm. 
\ja{Later in this article, the developed theoretical approach is applied to the experimental data obtained in \cite{kusyn2023}. Those experiments were done under the following conditions: The electrodes are in the sealed glass chamber evacuated to high vacuum (10$^{-5}$\,mbar), heated to 700\,K for several hours (baked out to reduce impurities) and then filled with argon gas of high purity of 99.9999\% to atmospheric pressure. 
In the previous work \cite{kusyn2023}, the emission in the pulsed barrier discharge was investigated by photomultipliers PMC-100-20 and PMC-100-4 and processed by time-correlated single photon counting module SPC-150 from Becker and Hickl GmbH. 
The voltage and current measurements were conducted by P6015A voltage probe (Tektronix) and current transformer CT-2 (Tektronix), respectively. 
Electrical characteristics were then captured by a high-definition oscilloscope Keysight DSO-2 204A. More details, parameters and a schematic of the experimental setup can be found in \cite{kusyn2023}.
}
 %\textcolor{red}{There isn't any figure \ref{model01}c showing the electrode arrangement at present.}
%The electrode arrangement for the 2D model is shown in Fig.\ \ref{geometry}b). 

\ja{
The time- and space-dependent fluid-Poisson model used in the present work to simulate the introduced pulsed barrier discharge 
\ja{is}
%has been 
described in \cite{becker2013,jovanovic2021}. 
Numerical calculations were performed for the given experimental conditions and different frequencies of the applied voltage.}
The applied voltage amplitudes were the same for all simulations as shown in Figure~\ref{geometry}(\ja{c} and \ja{(d)}), i.e.\ approx.\ 5.6\,kV for the nanosecond pulse. 
\ja{The rise time of the applied voltage pulse was about 40\,ns.}
The \ja{fluid-Poisson} model was solved in 1D (in further text denoted as 1D model) and 2D geometry (denoted as 2D model).
The 2D model considers the entire geometry shown in Figure~\ref{geometry}(a) \ja{and (b)}. The 1D modelling studies were performed along the discharge symmetry axis at $r$~=~0\,mm illustrated by the grey dashed line in Figure~\ref{geometry}(\ja{b}), neglecting radial effects. 

%As shown in Figure~\ref{geometry}(a), 
The voltage 
%(see Fig.\ \ref{model01}) 
was applied to the metal electrode (i.e.\ the electrode without the dielectric coverage, see the lower part %of the picture
in Figure~\ref{geometry}(a)) with a repetition frequency of  5, 10, 20, 50 and 100\,kHz, respectively. The other electrode was grounded. 
For repetitive discharges described by the  1D model, the modelling was performed until a quasi-periodic state was reached (typically ten periods). Typical current and voltage waveforms after reaching the periodic regime are shown in Figures~\ref{geometry}(\ja{c}) and (\ja{d}). 
\ja{Here, the memory voltage is the voltage drop across
the dielectric layer, which is determined by subtracting the gap voltage (voltage drop across the gas gap) from the applied voltage.} 
%In the 2D case, the modelling was performed until the streamer reached the cathode.
The 2D model was used to describe only the  very first discharge until the streamer reached the cathode. It was used to clarify the influence of radial effects and particularly the effect of line-of-sight integration
of measured light emission signals  (the radiative state densities) from the discharge. The model calculations in 2D  were limited to the first discharge due to the high computation cost including the full reaction kinetics scheme \ja{detailed in \cite{stankov2022}}.

%For the theoretical description of the spatiotemporal development of microdischarges, a time- and space-dependent hydrodynamic model was employed taking into account the spatially one-dimensional discharge geometry depicted in Fig.\ \ref{scheme}. The nonlinear system of partial differential equations comprises the continuity equations, the electron energy balance equation determining the spatiotemporal evolution of the electron energy density and Poisson's equation providing self-consistently the electric potential and the electric field (see for details in \cite{becker2013}). The particle fluxes and the electron energy flux in the x-direction are given in drift-diffusion approximation by \cite{sige1999}. The mean energy dependent transport coefficients for electrons as well as the rate coefficients of elastic collisions, exciting, de-exciting and ionizing collision processes have been obtained by solution of the stationary, spatially homogeneous electron Boltzmann equation in a convergent multi-term approach using a modified version of the method presented in \cite{leyh1998}.

%\textcolor{red}{Figure 1 is missing so far and Figure 2 does not show the waveform at present. }
% The 1D model was used to model a repetitive discharge, i.e.\ the sequence of the first and several further periods (usually ten), until the discharge reached a periodic regime. 

The fluid-Poisson model in 1D and 2D comprises the same set of particle balance equations, Poisson's equation and the electron energy balance equation. 
An additional balance equation for the surface charge density $\sigma$ was solved at the plasma-dielectric interface and used as boundary condition for the Poisson equation~\cite{jovanovic2023}.
More details about the  model implementation and the numerical procedure used to solve the equations can be found in \cite{jovanovic2021}.

\begin{table}[htb]
%    \extrarowheight2pt
    \caption{List of argon species considered in the 
    \ja{RKM}. %   model.%
    \label{species}}
    \begin{indented}
    \item[]\begin{tabular}{@{}rlr}\br
        Index & Species & Energy level [\unit[]{eV}] \\\mr% & Statistical weight\\\mr
        $1$ & $\mathrm{Ar[1p_0]}$ & 0 \\%& 1\\
        $2$ & $\mathrm{Ar[1s_5]}$ & 11.55 \\%& 5\\
        $3$ & $\mathrm{Ar[1s_4]}$ & 11.62 \\%& 3\\
        $4$ & $\mathrm{Ar[1s_3]}$ & 11.72 \\%& 1\\
        $5$ & $\mathrm{Ar[1s_2]}$ & 11.82 \\%& 3\\
        $6$ & $\mathrm{Ar[2p_{10}]}$ & 12.91 \\%& 24\\
        $7$ & $\mathrm{Ar[2p_9]}$ & 13.08 \\%& 24\\
        $8$ & $\mathrm{Ar[2p_8]}$ & 13.09 \\%& 24\\
        $9$ & $\mathrm{Ar[2p_7]}$ & 13.15 \\%& 24\\
        $10$ & $\mathrm{Ar[2p_6]}$ & 13.17 \\%& 24\\
        $11$ & $\mathrm{Ar[2p_5]}$ & 13.27 \\%& 24\\
        $12$ & $\mathrm{Ar[2p_4]}$ & 13.28 \\%& 24\\
        $13$ & $\mathrm{Ar[2p_3]}$ & 13.30 \\%& 24\\
        $14$ & $\mathrm{Ar[2p_2]}$ & 13.33 \\%& 24\\
        $15$ & $\mathrm{Ar[2p_1]}$ & 13.40 \\%& 24\\
        $16$ & $\mathrm{Ar^*[hl]}$ & 13.84 \\%& 122\\
        $17$ & $\mathrm{Ar^+}$ & 15.76 \\\mr%\\\mr
        $18$ & $\mathrm{Ar^*_2[^3\Sigma_u^+,v=0]}$ & 9.76\\
        $19$ & $\mathrm{Ar^*_2[^1\Sigma_u^+,v=0]}$ & 9.84\\
        $20$ & $\mathrm{Ar^*_2[^3\Sigma_u^+,v\gg0]}$ & 11.37\\
        $21$ & $\mathrm{Ar^*_2[^1\Sigma_u^+,v\gg0]}$ & 11.45\\
        $22$ & $\mathrm{Ar_2^+}$ & 14.50\\\br
    \end{tabular}
    \end{indented}
\end{table}

The 
reactions kinetics model (RKM)
considers the electron component, 22 heavy  particle species and about 400 collision and radiation 
\ja{processes.} 
%%%% processes as described in \cite{stankov2022}. 
The list of considered 
heavy particle species  
\ja{as well as their energy levels} 
is given in Table\,\ref{species}. Further details regarding the reaction kinetics are represented in~\cite{stankov2022}. 
This 
reaction kinetics model was designed 
\ja{for the analysis of the electrical characteristics and}  
for the description of production and loss channels of the %radiative and metastable 
excited species in 
\ja{gas discharge plasmas in the range from low to atmospheric pressure.} 
%%%the atmospheric-pressure discharge. 
The 2p states are here of special importance due to their dominant contribution to the optical emission spectrum. 
Each of the 2p states is described using the following equation:

%\ioponecol

%%% zde patri rovnice

\begin{eqnarray}
\frac{\mathrm{d}n_{2p_i}}{\mathrm{d}t}
&= \left[ n_e \bigg\{ n_{g}k_{g,2p_i}\bigg\} \right]_I+ \left[n_e \bigg\{\sum_{j=2}^{5} n_{1s_j}k_{1s_j,2p_i} \bigg\}\right]_{II} \nonumber\\
&+ \left[n_e \bigg\{ \sum_{m,m\neq i}^{} n_{2p_m,2p_i} k_{2p_m,2p_i} \bigg\}\right]_{III}  + \left[n_e \bigg\{ n_{hl}k_{hl,2p_i}\bigg\}\right]_{IV} + \left[n_e \bigg\{n_{Ar_2^+}k_{Ar_2^+,2p_i} \bigg\}\right]_{V}  \nonumber\\ 
&- \left[n_e n_{2p_i} \bigg\{ \sum_{j=2}^{5} k_{2p_i,1s_j}\bigg\}\right]_{VI} - \left[n_e n_{2p_i} \bigg\{ \sum_{m,m\neq i} k_{2p_i,2p_m} \bigg\}\right]_{VII} - \left[n_e n_{2p_i} \bigg\{ k_{2p_i,g} + k_{2p_i,hl}  + k_{2p_i,ion} \bigg\}\right]_{VIII}  \nonumber\\ 
&+ \left[n_g \sum_{m,m\neq i}^{} \bigg\{k_{2p_m,2p_i}\big(T_g\big)n_{2p_m} - k_{2p_i,2p_m}\big(T_g\big)n_{2p_i}\bigg\}\right]_{IX} - \left[n_g \sum_{j=2}^{5} \bigg\{k_{2p_i,1s_j}\big(T_g\big)n_{2p_i}\bigg\}\right]_{X}  \nonumber\\ 
&+ \left[n_{hl} A_{hl,2p_i}\right]_{XI} - \left[n_{2p_i} \sum_{j=2}^{5} \bigg\{ A_{2p_i,1s_j} \bigg\}\right]_{XII} 
\label{population}
\end{eqnarray}

\ioptwocol

\noindent
where $n_{2p_i}$ is the density of the given 2p state with $i$ ranging from 1 to 10, $n_{g}$ and $n_{e}$ are the ground gas ($\mathrm{Ar[1p_0]}$) and electron density, respectively, $E/N$ is the reduced electric field, $k_{g,2p_i}$ is rate coefficient for excitation of the $2p_i$ state from the ground state and similarly for other collision processes contributing to the population or depopulation of the respective $2p_i$ state. $A_{hl}$ denotes the Einstein coefficient of higher lying states $hl$ (lumped for all states energetically higher 
\ja{than the}  
2p$_1$ state, see \cite{stankov2022}), which are here lumped into one state. 
The rate equation includes gain processes from direct electron impact excitation 
from ground state (I), stepwise excitation from 1s states (II), excitation and de-excitation from other 2p states (III), de-excitation from higher lying state (IV) and electron-ion recombination (V). 
It further includes loss processes from electron impact de-excitation to 1s states (VI), excitation and de-excitation to other 2p states (VII) and to ground and higher lying states as well as stepwise ionization (VIII). 
\ja{Equation~(\ref{population})}
%Equation~\ref{population} 
also includes heavy-particle quenching processes from and to upper and lower 2p states (gain and loss) (IX), depopulation due to quenching processes to 1s states (X), population due to radiative de-excitation of the higher lying state (XI) and radiative de-excitation to 1s states (XII) described by the Einstein coefficients $A_{2p_i, 1s_j}$.
%electron quenching to $2p_x$, quenching of higher lying states by neutral atom collisions to $2p_x$, excitation by neutral to $2p_x$, collisional interaction between different $2p_x$ states or radiative deexcitation described by the Einstein coefficients $A_{2p_x, 1s_y}$ and others.
Referring to the last term in the equation, this \ja{RKM} allows \ja{to monitor }the densities of relevant radiative excited species responsible for the strong lines in argon spectra during the discharge, as observed in experiments \cite{kusyn2023,zhu2010,simek2018}. 
%The use of this sophisticated model is therefore the requirement for possible determination of the electric  field using method of line ratios. 
%In addition to volume processes, the model employs the same boundary conditions and corresponding surface coefficients as in \cite{becker2013}.

To allow proper comparison of the very first discharge events for the 1D and 2D model, the same initial number density of particle species and mean electron energy were set in both cases. 
Namely, the initial density of excited argon atoms and molecules and positive argon ions was set to 10$^9$\,m$^{-3}$ and the initial density of electrons was set to 2$\times 10^9$\,m$^{-3}$ to assure quasi-neutral initial conditions. 
An initial mean electron energy of 1.5\,eV was assumed in the gap and the gas temperature was fixed at 300\,K for all model calculations. 

\section{Results}
\label{Results}

%\DLR{Remark DL: Some introductory text is still required here to first indroduce the results of Figure~\ref{model01} and explain the stucture of this section.}

In this section, we present the results of numerical simulations, first for the 2D geometry where we describe the data treatment for investigation of the effect of radial or axial signal integration. 
Second, for the 1D geometry where the discharge dynamics for different frequencies is presented together with quantitative description of all parameters ($E/N$, streamer velocity, density of electrons, metastables and 2p states) important for development and validation of the CRM for the intensity ratio method. 
Further, the simple CRM for the intensity ratio method is described together with all its components and with care to evaluation of the effective lifetimes of the utilized 2p states. 
Finally, the semi-automated method is presented and used to evaluate the data from 1D simulations for all investigated frequencies and a comparison with the experimental results is made.

\begin{figure*}[!t]
\centering
%a)
%\includegraphics[clip = true,width=0.45\columnwidth]{voltages.png}
%b)
%\includegraphics[clip = true,width=0.4\columnwidth]{model01.png}
%\includegraphics[scale=1]{Figure1a.pdf}
%\includegraphics[scale=1]{Figure1b.pdf}
\includegraphics[scale=1]{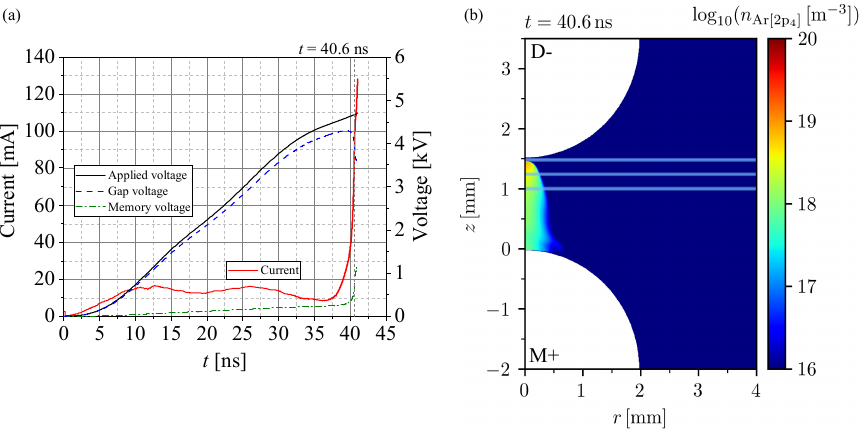}
%\end{center}
\caption{Illustration of (a) the temporal evolution of the current and voltage and (b) the spatial distribution of 2p$_4$ state density at a given time calculated using the 2D model. The metal electrode and the dielectric surface are denoted as M and D, respectively, where `$+$'?? sign indicates the temporary anode and `$-$'?? the temporary cathode. The moment of the streamer arrival at the cathode at $t$ = 40.6\,ns was taken for an illustration of the streamer spatial profile. The three shown horizontal lines at $z$ = 1.0, 1.24 and 1.48\,mm highlight the positions for more detailed investigations and for comparison with measured data from \cite{kusyn2023}.}
\label{model01}
\end{figure*}

As a brief introduction \ja{to the dynamics of} the nanosecond pulsed barrier discharge %, to show the distribution of excited argon states after the breakdown 
and to illustrate the sampling procedure, %for the analysis of the detailed reaction kinetics, 
the results of the 2D modelling are shown first. 
Figure~\ref{model01} shows the voltage and current waveforms together with the spatial \ja{distribution of the 2p$_{\mathrm{4}}$ state density in} the streamer at the time of 40.6\,ns calculated 
\ja{by means of} 
%with 
the 2D model for the first discharge event. 
\ja{That is the calculations start at $t=0$. The time instant of 40.6\,ns is highlited by a dashed vertical line also in Fig.~\ref{model01}(a).}
%In Fig.\ \ref{model01}b), the propagation of the streamer is shown at the time of 40.6\,ns and
\ja{The discharge starts with increased current to the level of about 15\,mA. This is due to the displacement current and initial electron multiplication in the gas gap. After the local free charge density crosses a certain threshold, the streamer starts and propagates towards the cathode. This is reflected by a rapid increase of the discharge current. 
In this work, we are interested in this initial phase of streamer propagation. 
After the streamer impacts onto the cathode the transient glow discharge plasma is created and with the accumulation of surface charges on the dielectrics the discharge is quenched. 
The discharge development after the streamer impact onto the dielectric was studied in detail for the same configuration in a sine-driven single-filament atmospheric-pressure DBD in \cite{jovanovic2022,jovanovic2023}.}

Three horizontal lines drawn in Figure~\ref{model01}(b) 
%\textcolor{red}{There isn't any figure \ref{model01}c showing the propagation of the streamer.}
%These lines 
denote positions which were selected to %theoretically study 
analyse 
the detailed reaction kinetics, dominantly of the 2p states. Experimentally, the spectra originating from 2p radiative states were measured for nanosecond pulsed discharge at the selected positions of 0.44, 0.88 and 1.24\,mm as in \cite{kusyn2023}. 
%In the present article, 
Here, additional positions were selected at 1.0 and 1.48\,mm. 
%A more detailed analysis of the discharges simulated with 1D and 2D models is given in the following section.

%\begin{figure}[h!]
%\begin{center}
%%a)
%%\includegraphics[clip = true,width=0.45\columnwidth]{voltages.png}
%%b)
%%\includegraphics[clip = true,width=0.4\columnwidth]{model01.png}
%\includegraphics[scale=0.49]{Electrodes_Geometry1.png}
%\includegraphics[scale=0.49]{Electrodes_Geometry2.png}
%\end{center}
%\end{figure}

%determine number of excited particles in this volume. 
%Note that the volume integration was performed, see further.

The results %of 
%both numerical models
of the 1D and 2D model 
were sampled for post-processing using 
% high spatial resolution (2\,$\mu$m)
a spatial resolution of 2\,$\mu$m 
along the discharge axis and a temporal resolution of $20\,\mathrm{ps}$. From this data set, the evaluation was performed at %desired 
the positions 
%1.0\,mm, 1.24\,mm and 1.48\,mm). 
$z= 0.88, 1.00$, 1.24 and 1.48\,mm. 
The data evaluated from a single point 
%(2\,$\mu$m wide) 
of the discharge axis ($r = 0$\,mm) are described as ``point data" 
in the following. %further in the article.
%If used in data evaluation, this precision is described as ``point data". % in 1-D and 2-D model. 
%These values were used as the reference for comparison of the two methods. 
%For the comparison with the experiment and for results close to the experimental possibilities, the axial averaging over 30\,$\mu$m in the $z$ axis (described as ``averaged data") as well as radial integrating (described as ``integrated data") of the densities (i.e.\ signals for the OES) were performed, similarly to the experiment (see e.g.\ \cite{jansky2021,kusyn2023}). 

%Electric field, electron, 2p state and other selected particle densities were sampled at every 100\,$\mu$m on the axis, similarly to the experiment (see e.g.\ \cite{jansky2021,kusyn2023}). 

In the experiment, the discharge filament is typically projected perpendicularly onto the monochromator slit, which is then by its opening (usually few tens of microns) defining the axial resolution of the measurement, typically 30\,$\mu$m. 
%of the spatial distribution of the emission intensity of the discharge. 
%covers the discharge in radial direction and acquires all the emitted light during the specified time. 
The 1D model does not deliver the radial structure of the discharge. 
%, only axial values were used for the analysis---for 2\,$\mu$m resolution (point data, i.e.\ from a single cell at the axis) as well as for axially integrated over 30\,$\mu$m (summation of state densities over 15 cells on the axis). 
In addition to  point data, results of particle number densities axially integrated over 
30\,$\mu$m (summation over 15 cells on the discharge axis) are shown. 
%For the 2D model, however, in addition to the axial ``point" and ``axially integrated" values, the number densities have been integrated in axial (30\,$\mu$m) and radial direction as well (described as integrated data later), 
Moreover, the particle number densities have been integrated in axial 
%(30\,$\mu$m) 
and radial direction using the 2D modelling results  
to reproduce the complete signal input to the detector as in the experiment.  
%(see Figure~\ref{model01}). 
These results are referred to as ``integrated data" in the following. 
The integration of the particle number  densities obtained from the 2D model was done using the equation

\begin{equation}
\label{tau}
n_{p,int} = 2\pi \int_{0}^{R_d} \int_{z_1}^{z_2} n_p\,r\, \mathrm{d}r\,\mathrm{d}z\,,
\end{equation}

\noindent
where $n_{p,int}$ are integrated densities $n_p$ of species $p$, $z_1$ and $z_2$ are coordinates at the $z$-axis, where $\Delta z = z_2 - z_1 = 30\,\mu$m, and $R_d$ = 4\,mm is the radius of the computational domain. 
%As mentioned above, besides the integrated data, the average value of the electric field in the 30\,$\mu$m long axial region in 1D and 2D model was also used for the comparison.

In following subsections, we investigate the discharge dynamics for different frequencies and the 2p states suitable for the intensity ratio method.

\begin{figure*}[!t]
\centering
%a)
%\includegraphics[clip = true,width=0.95\columnwidth]{Iexp1.png}
%b)
%\includegraphics[clip = true,width=0.95\columnwidth]{Eexp1.png}
%c)
%\includegraphics[clip = true,width=0.95\columnwidth]{Dexp1.png}
\includegraphics[scale=1]{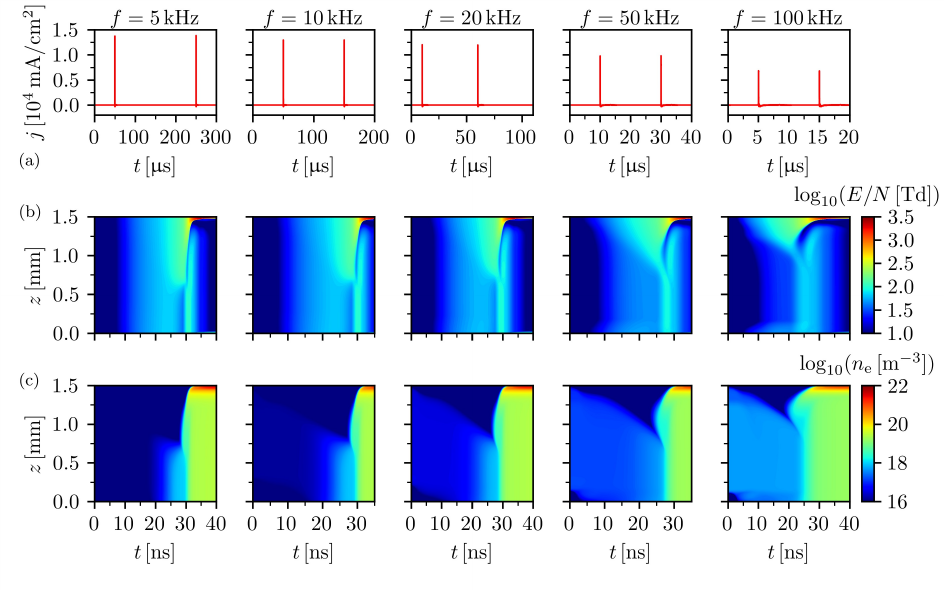}
%\end{center}
\caption{1D modelling results for nanosecond pulsed discharges: (a) period evolution of the current density $j$ as well as  
(b) magnitude $|E|/N$ of the reduced electric field and (c) electron density $n_\mathrm{e}$
%Electrical current densities as result from the 1D model simulations for nanosecond pulsed discharge a), electrical field b) and electron densities c) 
during the streamer propagation and arrival to the dielectric surface for the same conditions. The cathode is placed under the dielectric surface as described earlier.  Results for frequencies in the range from 5 to 100\,kHz are presented. Note that the time after reaching quasi-periodic condition was subtracted in the figures.
}
\label{model04}
\end{figure*}

\subsection{Discharge dynamics for different  frequencies}\label{sec:different_frequencies}

The asymmetric barrier discharge in atmospheric pressure argon was simulated for frequencies ranging from 5 to 100\,kHz using the 1D model. 
%The result of frequency variation simulations 
Modelling results of that frequency variation 
are shown in Figure~\ref{model04}, 
where the current densities $j$ (a), magnitudes $|E|/N$ of the reduced electric field  (b) and electron number densities $n_\mathrm{e}$ (c) are presented. 
%The single discharge (first breakdown event in the sequence) conditions are discussed later. 

Apparently the current density peaks at the rising slope of the nanosecond voltage pulse,  where a strong discharge takes place. 
It is apparent 
from Figure~\ref{model04}(a) 
that the maximum current density decreases with increasing frequency. 
Additionally, discharge with a significantly smaller peak current density and of considerably longer duration takes place on the falling slope of the applied voltage pulse\ja{, see Figure~\ref{geometry}.}  %(not highlighted as it is not in the focus of this work). 
Such an event is a result of the discharging of the residual charge left on the dielectric surface during the rising slope (see also \cite{hoft2020}). 
Note that this is an asymmetric barrier discharge and thus also the secondary electron emission coefficient or the local influence of the accumulated surface charge are different for the respective 
\ja{surface material (metal or dielectric)} 
%%%electrode material 
at a given polarity.  
In this study we focus 
%%ourselves 
on %the positive half-period, i.e.\ 
the rising slope of the applied voltage, when the metal electrode is anode and the streamer propagates towards the cathode covered by 
\ja{a dielectric,} 
%dielectrics, 
as shown in Figure~\ref{model01}(b). 
The current density maxima at rising slope decrease from 1.4$\times 10^{5}$\,A/m$^2$ for 5\,kHz to almost a half at 100\,kHz. 
As the current density decreases for the rising slope for higher frequencies, one can see also the smaller and longer  current density hump for the discharges at the falling slope.

The corresponding development of $|E|/N$  in the barrier discharges is shown in Figure~\ref{model04}(b). 
%, corresponding to the same frequencies as the current densities above. 
Here, only the important time interval of the half-period is selected, i.e.\ where the streamer manifests itself with an enhanced reduced electric field strength as it propagates towards the cathode and the electric field peaks at the dielectric surface vicinity. 
The maximum values of the corresponding $E/N$ scales for each 
%sub-figure 
condition 
clearly show that 
%also 
the peak $E/N$ 
%is decreasing with the 
decreases with increasing 
frequency. The reduced electric field strength peaks as the streamer reaches the dielectric surface and for 5\,kHz it is approx.\ 2300\,Td. 
The peak value for the 100\,kHz case is approx.\ 1840\,Td. 
Corresponding results and 
maximum values of the electron density  
in the middle of the gap ($z = 0.75$\,mm) and in the cathode region are summarized in~Table~\ref{MaxDensities}. 

From the local maxima of the $E/N$ spatiotemporal distribution, the moving streamer head, one can determine the velocity of the propagating streamer $v_\mathrm{streamer}$. 
Here, %only 
the mean velocity was determined \ja{from the time of the movement of the electric field maximum between two points near the cathode at 1.2 and 1.4 mm.}
%. This was done for given frequencies 
The results are also given 
%also 
in Table~\ref{MaxDensities}. It is apparent that 
$v_\mathrm{streamer}$
%the velocity 
decreases with increasing frequency. 
%These data are presented in Table~\ref{MaxDensities}, together with the maximum data for electron densities, also for the first discharge (i.e.\ discharge without enhanced preionization). Values for the position of 0.75\,mm (the middle of the gap) are presented as well.

\begin{table*}[htb]
\caption{Maxima of the electron number density and reduced electric field in the middle of the gap ($z = 0.75$\,mm) and in the cathode region 
\ja{(CR)} 
for first and quasi-periodic discharge. 
\ja{The last} 
%Last 
row displays the streamer velocity determined from the movement of the electric field maximum between $z = 1.2$\,mm and $z = 1.4$\,mm during streamer propagation.
\label{MaxDensities}}
\begin{tabular}{@{}l*{15}{l}||}\br
& Quantity & $1^\mathrm{st}$ disch.  & 5\,kHz & 10\,kHz & 20\,kHz & 50\,kHz & 100\,kHz \\\mr
& $n_\mathrm{e}( 0.75 \, \mathrm{mm})  \, \mathrm{[10^{19} \, m^{-3}]}$ & $3.65$ & $3.10$ & $2.30$ & $2.87$ & $1.87$ & $1.92$ \\
& $E/N(0.75 \, \mathrm{mm}) \, \mathrm{[Td]}$ & $168$ & $121$ & $103$ & $86$ & $82$ & $68$\\\mr
& $n_\mathrm{e}\ja{(\mathrm{CR})}\, \mathrm{[10^{21} \, m^{-3}]}$ & $3.87$ & $2.33$ & $1.10 $ & $1.85$ & $1.05$ & $0.84$ \\
& $E/N\ja{(\mathrm{CR})} \, \mathrm{[Td]}$ & $2702$ & $2297$ & $2241$ & $2192$ & $2065$ & $1842$\\\mr
& $v_\mathrm{streamer} \, \mathrm{[10^6\,m/s]}$ & $0.28$ & $0.21$ & $0.18$ & $0.15$ & $0.07$ & $0.05$\\\mr
\end{tabular}
\end{table*}

\ja{The corresponding development of the electron number density 
for the relevant time intervals of the half-period of} 
%The electron density development for the crucial time intervals in 
the discharges is presented in Figure~\ref{model04}(c) for the respective frequencies. 
Apparently, the higher the frequency the higher is the residual electron density in the gap which influences the initial ionization and the start of the streamer. 
The initial electron densities, describing the preionization before the discharge ignites, 
%initiates, 
are given 
in Table~\ref{AvgDensities}. They 
range from approx.\ 10$^{15}$\,m$^{-3}$ for 5\,kHz to 10$^{17}$\,m$^{-3}$ for 100\,kHz. 
%(cf table~\ref{AvgDensities}). 
At the same time, it is apparent that the maximum electron density in the established transient discharge channel decreases with the increase of the frequency, which is a result of the lower electric field in the gap. 
%as described above. 
The electron density in the discharge channel decreases from %approx.\ 
$3.1 \times 10^{19}$\,m$^{-3}$ for 5\,kHz to %approx.\ 
$1.92  \times 10^{19}$\,m$^{-3}$ for 100\,kHz. (cf.\ Table~\ref{MaxDensities}). 
%The exact values are also presented in the table \ref{MaxDensities}.
%\textcolor{red}{The values in Table~\ref{MaxDensities}) at $z=0.75$\,mm differ from the values here in the text!}

The above parameters of the investigated microdischarges are strongly influenced by the preceding discharges, i.e.\ the initial conditions at the beginning of each period in the periodic state. These initial conditions of interest %\st{in this article} 
are given by the number densities of electrons, metastables and other excited states (2p and 1s) in the gap. Their values for each frequency are given 
in Table~\ref{AvgDensities}. Such data serves as reference for further discussion of the applicability of the proposed diagnostic method using the densities of the 2p states.

\begin{table*}[htb]
%    \extrarowheight2pt
\caption{Initial number densities of electrons and excited atomic argon species at the beginning of the period (i.e. $t = T$) in $\mathrm{m}^{-3}$ for different repetition frequencies (for first and quasi-periodic discharge) averaged %along the middle of the gap (
between $z = 0.5$ and $z = 1$\,mm, %),
far from the sheath regions.
\label{AvgDensities}}
%\begin{indented}
\begin{tabular}{@{}l*{15}{l}||}\br
%\item[]\begin{tabular}{@{}rlr}\br
& Species & $1^{st}$ disch.  & 5\,kHz & 10\,kHz & 20\,kHz & 50\,kHz & 100\,kHz \\\mr
 & $\mathrm{e}$ & $2 \times 10^9$ & $6.59 \times 10^{15}$ & $1.45 \times 10^{16}$ & $3.46 \times 10^{16}$ & $1.47 \times 10^{17}$ & $4.73 \times 10^{17}$ \\
 & $\mathrm{Ar[1s_5]}$ & $10^9$ & $1.98 \times 10^{12}$ & $9.05 \times 10^{12}$ & $5.16 \times 10^{13}$ & $7.61 \times 10^{14}$ & $5.38 \times 10^{15}$ \\
 & $\mathrm{Ar[1s_4]}$ & $10^9$ & $1.81 \times 10^{13}$ & $8.29 \times 10^{13}$ & $4.70 \times 10^{14}$ & $6.61 \times 10^{15}$ & $4.17 \times 10^{16}$ \\
 & $\mathrm{Ar[1s_3]}$ & $10^9$ & $3.19 \times 10^{12}$ & $1.45 \times 10^{13}$ & $8.13 \times 10^{13}$ & $1.12 \times 10^{15}$ & $7.00 \times 10^{15}$ \\
 & $\mathrm{Ar[1s_2]}$ & $10^9$ & $3.44 \times 10^{12}$ & $1.56 \times 10^{13}$ & $8.75 \times 10^{13}$ & $1.19 \times 10^{15}$ & $7.56 \times 10^{15}$ \\
 & $\mathrm{Ar[2p_{10}]}$ & $10^9$ & $1.94 \times 10^{9}$ & $8.78 \times 10^{9}$ & $4.88 \times 10^{10}$ & $6.57 \times 10^{11}$ & $3.98 \times 10^{12}$ \\
 & $\mathrm{Ar[2p_9]}$ & $10^9$ & $1.57 \times 10^{9}$ & $7.10 \times 10^{9}$ & $3.94 \times 10^{10}$ & $5.32 \times 10^{11}$ & $3.22 \times 10^{12}$ \\
 & $\mathrm{Ar[2p_8]}$ & $10^9$ & $2.08 \times 10^{9}$ & $9.42 \times 10^{9}$ & $5.23 \times 10^{10}$ & $7.05 \times 10^{11}$ & $4.27 \times 10^{12}$ \\
 & $\mathrm{Ar[2p_7]}$ &$10^9$ & $7.51 \times 10^{8}$ & $3.40 \times 10^{9}$ & $1.89 \times 10^{10}$ & $2.54 \times 10^{11}$ & $1.54 \times 10^{12}$ \\
 & $\mathrm{Ar[2p_6]}$ &$10^9$ & $5.68 \times 10^{9}$ & $2.57 \times 10^{10}$ & $1.43 \times 10^{11}$ & $1.93 \times 10^{12}$ & $1.16 \times 10^{13}$ \\
 & $\mathrm{Ar[2p_5]}$ &$10^9$ & $1.30 \times 10^{9}$ & $5.89 \times 10^{9}$ & $3.27 \times 10^{10}$ & $4.41 \times 10^{11}$ & $2.67 \times 10^{12}$ \\
 & $\mathrm{Ar[2p_4]}$ &$10^9$ & $7.86 \times 10^{8}$ & $3.55 \times 10^{9}$ & $1.97 \times 10^{10}$ & $2.66 \times 10^{11}$ & $1.61 \times 10^{12}$ \\
 & $\mathrm{Ar[2p_3]}$ &$10^9$ & $4.41 \times 10^{8}$ & $1.99 \times 10^{9}$ & $1.11 \times 10^{10}$ & $1.49 \times 10^{11}$ & $9.03 \times 10^{11}$ \\
 & $\mathrm{Ar[2p_2]}$ & $10^9$ & $5.59 \times 10^{8}$ & $2.53 \times 10^{9}$ & $1.41 \times 10^{10}$ & $1.89 \times 10^{11}$ & $1.15 \times 10^{12}$ \\
 & $\mathrm{Ar[2p_1]}$ &$10^9$ & $1.72 \times 10^{9}$ & $7.80 \times 10^{9}$ & $4.33 \times 10^{10}$ & $5.84 \times 10^{11}$ & $3.53 \times 10^{12}$ \\\mr
    \end{tabular}
    %\end{indented}
\end{table*}

\begin{figure*}[!t]
\centering
%\begin{figure}[h!]
%\begin{center}
%a)
%\includegraphics[clip = true,width=0.95\columnwidth]{fields.png}
%b)
%\includegraphics[clip = true,width=0.45\columnwidth]{2D_overview_logscale.png}
%c)
%\includegraphics[clip = true,width=0.45\columnwidth]{2D_overview_logscale.png}
\includegraphics[scale=1]{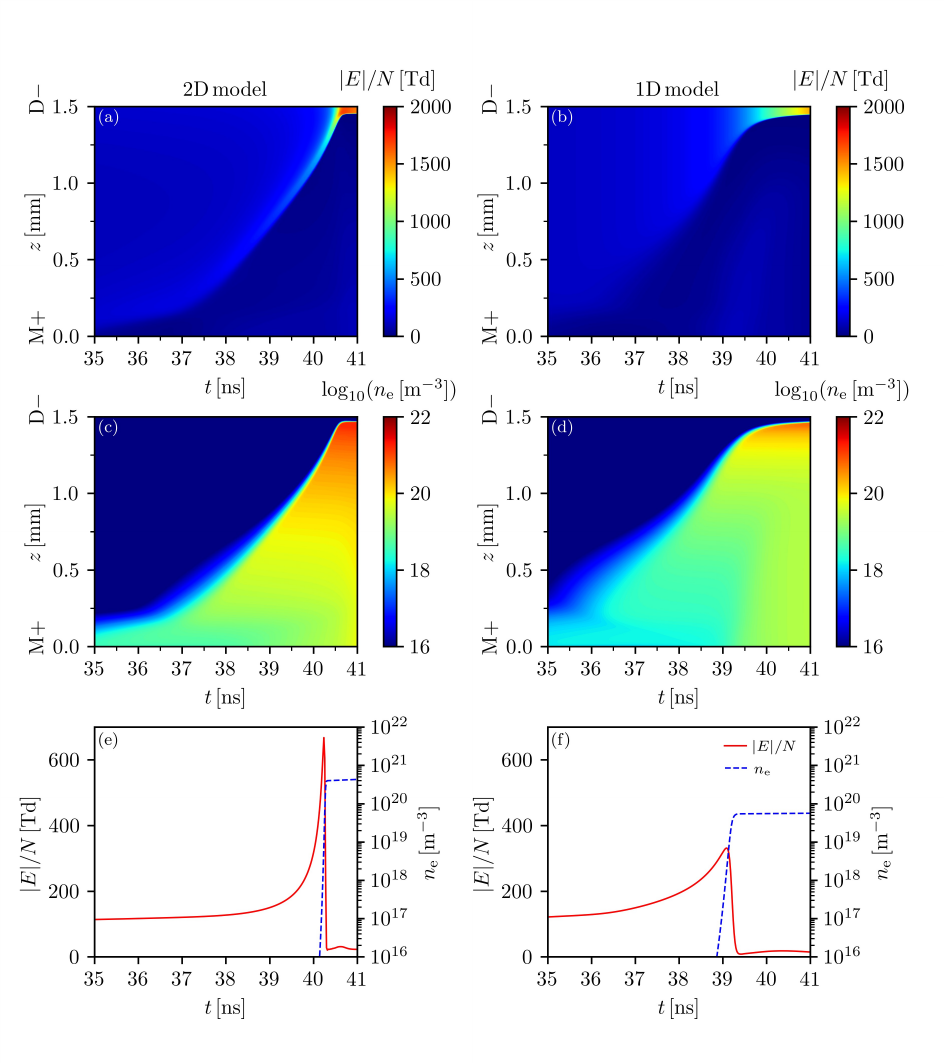}
%\end{center}
\caption{
The magnitude of the reduced electric field and electron number density obtained by (a) and (c) 2D and (b) and (d) 1D  model calculations. The development of these quantities is also highlighted for the position 1.24\,mm in (e) and (f). %\ja{I have uploaded the updated Figure 3 (file Figure3.pdf) with the spatiotemporal evolution of electron number density and electric field. The evolution of the field at 1.24\,mm was not shown since I was not sure how to present it properly. Maybe we could add the evolution of the reduced electric field at three positions as the additional panels d) and e)?} 
}
\label{model02}
\end{figure*}

The 2D 
%simulation was included in this work 
modelling studies serve 
to investigate the effect of radial %and axial 
signal integration as it takes place during measurements. Barrier discharge filaments at atmospheric pressure usually show a diameter of %few 
tens to hundreds of microns and the possibilities to resolve the radial structure are therefore very limited, if not impossible (compare \cite{simek2012}). 
\ja{As the given fluid model is solved both in spatially 1D and 2D geometries, it is important to point out some differences in their results, which are caused by neglecting radial effects in the 1D model.} 
The comparison of the first discharge (single shot) simulation using the 2D and the 1D model is shown in Figure~\ref{model02}. 
%Apparently, the 2D model catches the streamer development with better precision than in the case of 1D model. 
Both cases show a gradual increase of the electron density with increasing applied voltage, leading to the accumulation of  space charges near the
temporary anode. The accumulated space charge leads to the inception of the positive streamer near $z \approx 0.25 \,\mathrm{mm}$ %at 
around $t = 37 \,\mathrm{ns}$ in the 2D model and $36.5 \,\mathrm{ns}$ in the 1D model. As the streamer propagates towards the cathode,
$|E|/N$ and the electron number density start to increase. Note that they increase more gradually in the 1D model due to  
\ja{the neglect of}
%not taking into account 
the radial constriction of the channel, which results in lower $|E|/N$ in comparison to the 2D model. This is reflected 
\ja{by} 
%in %The 2D model describes %also the $E/N$ peak in the streamer head with 
a much more rapid increase to the maximum in the 2D model in comparison to the 1D model. 
%, see Figure~\ref{model02}. %This clear distinction of the peak $E/N$ is missing in the 1D approach.
At the same time, the acceleration of the streamer towards the dielectric-covered cathode is well visible in both cases and is stopped first directly in the vicinity of the dielectric surface. This is in agreement with the experimental data on streamers in pulsed or sinusoidal barrier discharges in 
argon~\cite{kusyn2023,kloc2010}. 
%argon, compare, e.g.\ \cite{kusyn2023,kloc2010}. 
Note that in contrast to the 1D results, where the streamer slows down much earlier (around $t = 39.7 \,\mathrm{ns}$) and farther 
away 
from the cathode (hundreds of microns), %and the $E/N$ peak is lower, 
in 2D model 
it approaches the cathode to a few tens of microns  around $t = 40.6 \,\mathrm{ns}$ resulting in a stronger $|E|/N$. 
%Limited description of the electric field in the close vicinity of the dielectric surface and in the streamer head using the 1D model maybe therefore one of the reasons of increased uncertainty for the intensity ratio evaluation of $E/N$ in this area, see further in the text. 
However, the electron number density in the 1D model keeps increasing over time in the cathode region, eventually reaching a similar order of magnitude as in the 2D model (cf.\ Table~\ref{MaxDensities}).

\ja{It is important to note that the primary objective of this work is not to determine the electric field as accurately as possible or to validate the accuracy of the models against the experiment. 
Rather, the aim is to employ the modelling results to evaluate the line intensity ratio method for argon and find the most suitable combination of states for electric field determination, in general.  Since both the 1D and the 2D fluid-Poisson model self-consistently couple the electric field with the species densities, this can be achieved by using the computationally more efficient 1D model. 
%\ja{HERE NEEDS TO BE DISCUSSED IN RELATION TO EQUATION 1 !!!}
%, this evaluation can be achieved even 
%\ja{by} 
%using the 
%\ja{\st{less accurate}}  
%1D model.}
%\DLR{Remark DL: It is little helpful to declare the 1D model as less accurate!
}

As mentioned earlier, semi-automated evaluation of a large amount of complex %simulations 
model calculations 
is time consuming and we are therefore limited here to the 1D modelling results. 
Within all 1D and 2D simulations, the distributions of densities of all other species were obtained, too. 
Such data are 
%then 
analysed later in the text together with the effect of density (signal intensity in experiment) integration on the intensity ratio for $E/N$ determination.

\subsection{Investigation of 2p states suitable for the intensity ratio method}\label{sec:semiautomatedprocedure}

The intensity ratio method originates from the simplification of the full reaction kinetics model (RKM, as described by the equation\,(\ref{population})) to a very few processes which dominate the populations of the investigated radiative states\ja{, the so called simple CRM}. For the case of air, as shown in Obrusník {\it et al.}\,\cite{obrusnik2018}, these processes are the direct electron impact excitation from the ground state as the only gain processes, the spontaneous emission, given by the radiative lifetime, and the collisional quenching by collisions with neutral gas. 
Here for the case of argon plasma, due to the presence of argon metastable states and other processes as denoted in the equation\,(\ref{population}), such simplification is expected to be problematic.
We have approached this issue in the following way: all processes (except the direct electron impact excitation from the ground state of argon) are lumped into an effective lifetime parameter, which is potentially time- and space-dependent (as basically all the source-term processes given in the equation\,(\ref{population}) are). 
We then investigate the possible quantification of this effective lifetime parameter of a given 2p state best suitable for experimental use. 
With this, the balance equation reads

\begin{equation}
\label{population2}
%\begin{split}
%\begin{align}
\frac{\mathrm{d}n_{2p_i}(z,t)}{\mathrm{d}t} = n_e(z,t) \cdot n_{g} \cdot k_{g,2p_i}(E(z,t)/N) - \frac{n_{2p_i}(z,t)}{\tau_{\mathrm{eff}}^{2p_i}(z,t)},  %\bigg\}
%\end{split}
%\end{align}
\end{equation}

\noindent
where $z$ describes the axial coordinate in the gas gap and $\tau_{\mathrm{eff}}^{2p_i}$ denotes the 
\ja{respective} 
effective lifetime. We assume the ground state concentration to be constant in the gap. 
%We can then continue with the derivation of an equation for the intensity (in our case the density) ratio method in the following way. We 
To derive an equation for the intensity (in our case the density) ratio method we 
select two 2p states $i$ and $j$ and divide their balance 
%equation. With small rearrangement this results in the following relation:
equations yielding the relation 

\begin{equation}
\label{ratioeq}
%\begin{split}
%\begin{align}
\frac{
%\frac{
\mathrm{d}n_{2p_i}(z,t)
/ %}{
\mathrm{d}t
%} 
+ 
%\frac{
n_{2p_i}(z,t)
/ %}{
\tau_{\mathrm{eff}}^{2p_i}(z,t)
%}
}{
%\frac{
\mathrm{d}n_{2p_j}(z,t)
/ %}{
\mathrm{d}t
%} 
+ 
%\frac{
n_{2p_j}(z,t)
/  %}{
\tau_{\mathrm{eff}}^{2p_j}(z,t)
%}
}
= 
\frac{k_{g,2p_i}(E(z,t)/N)}{k_{g,2p_j}(E(z,t)/N)}.  %\bigg\}
%\end{split}
%\end{align}
\end{equation}

% \begin{figure}[h!]
% \begin{center}
% a)
% \includegraphics[clip = true,width=0.45\columnwidth]{argon12.png}
% b)
% \includegraphics[clip = true,width=0.45\columnwidth]{rate_coeficients_ratio_2p2.png}
% c)
% \includegraphics[clip = true,width=0.45\columnwidth]{rate_coeficients_ratio_2p3.png}
% d)
% \includegraphics[clip = true,width=0.45\columnwidth]{rate_coeficients_ratio_2p4.png}
% \end{center}
% \caption{
% rates a) and rate ratios for upper 2p$_2$ b), 2p$_3$ c) and 2p$_4$ d) states 
% }
% \label{ratios}
% \end{figure}

\begin{figure*}%[!t]
\centering
\includegraphics[clip = true,width=0.89\columnwidth]{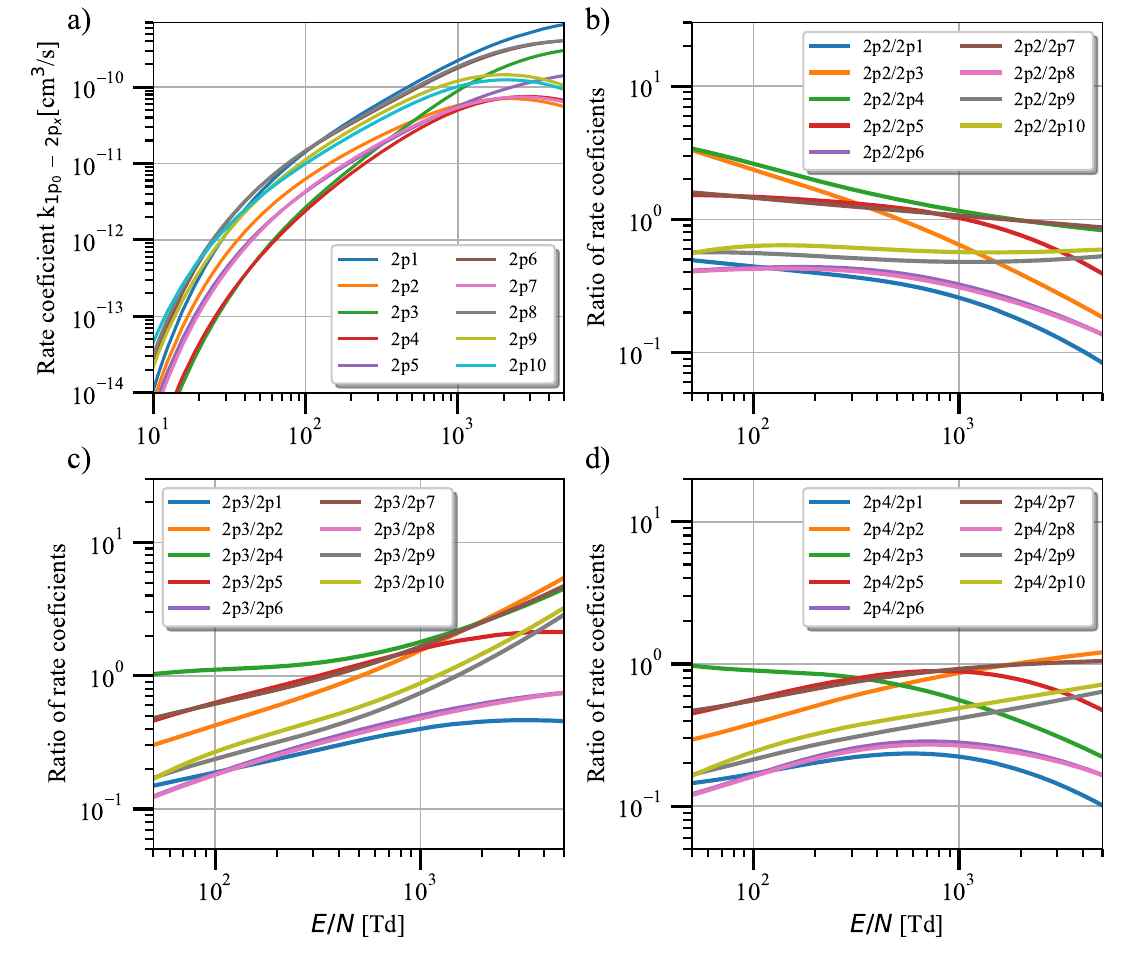}
\caption{
Rate coefficients of 2p$_{\textrm{1-10}}$ argon states (a) and ratios of rate coefficients with upper state 2p$_{\textrm{2}}$ (b), 2p$_{\textrm{3}}$ (c) and 2p$_{\textrm{4}}$ (d) as a function of $E/N$. 
}
\label{ratios}
\end{figure*}

Having the spatiotemporal development of the 2p state densities from the 
\ja{1D model simulations,} 
%%%model, 
%\DLR{Remark DL: Is that what you mean}
we rely only on the rate coefficient ratio for the selected pair of 2p states and on the variable $\tau_{\mathrm{eff}}^{2p_i}(z,t)$. 
The rate coefficients for electron collision processes were obtained by solving the electron Boltzmann equation in multi-term approximation \cite{leyh1998} using the cross-section set from \cite{zatsa2014,zatsa2018} as presented in \cite{stankov2022}. The rate coefficients and their selected ratios are presented in Figure~\ref{ratios}. 
Dominantly, the ratios including the rate coefficients for direct excitation of the 2p$_2$, 2p$_3$ and 2p$_4$ states 
%will be 
are 
discussed here, see \cite{kusyn2023} and text below. 
The proper determination of the effective lifetimes $\tau_{\mathrm{eff}}^{2p_i}(z,t)$ is more challenging and requires certain assumptions.
As all the processes determining the effective lifetimes are included in the fluid-Poisson  model, we 
%have 
selected the following three ways for their quantification considering that a quantification or use of the lifetimes from or for measured data is the ultimate goal. 

First, we include only all loss processes $S_{{2p_i}, \textrm{ all losses}}(z, t)$ in the $\tau_{\mathrm{eff}}^{2p_i}(z,t)$ for given 2p$_i$ and denote it as type 1 effective lifetime $\tau_{\mathrm{eff,1}}^{2p_i}(z,t)$ defined as %follows:

\begin{equation}
\label{tau1eq}
\tau_{\mathrm{eff,1}}^{2p_i}(z,t) = \frac{1}{S_{{2p_i}, \textrm{ all losses}}(z, t)}.
\end{equation}

\begin{figure}%[htb]
\centering
\includegraphics[clip = true,width=0.95\columnwidth]{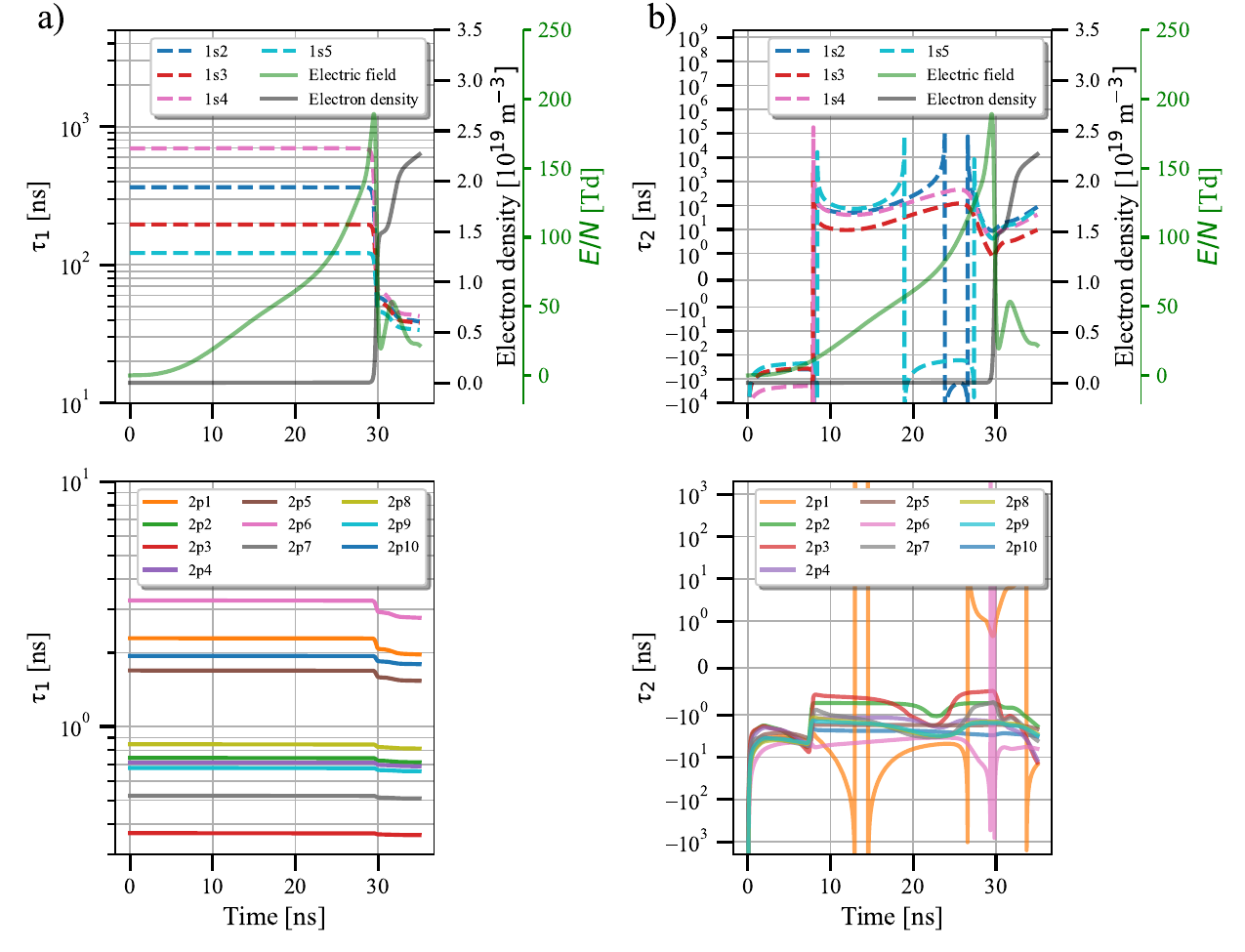}
\caption{
Temporal development of $\tau_{\mathrm{eff,1}}(z,t)$, (a), and $\tau_{\mathrm{eff,2}}(z,t)$, (b), \ja{during the quasi-periodic discharge} \ja{as obtained from 1D simulations}. The first row corresponds to the effective lifetimes of 1s$_{2-5}$ states compared with the $E/N$ and electron number density waveforms. The second row depicts effective lifetimes of 2p$_{1-10}$ states. The effective lifetimes are shown for a discharge frequency of 10\,kHz and spatial coordinate of 1.24\,mm. Note that in figures (b) the interval -1\,:\,1 is linear for better data presentation.}
%\textcolor{blue}{I have included some alternative images in the overleaf files Figure5.pdf (absolute values) and Figure5b.pdf (negative log scale)}
%}

\label{taus}
\end{figure}

\noindent
Such %a 
choice has its importance and applicability in the relative spatiotemporal stability of 
this 
lifetime and give us 
an %therefore 
almost %a 
constant number without large spatiotemporal variations. This was evaluated from the %simulation 
\ja{1D simulations} 
of the repetitive nanosecond pulsed barrier discharge under consideration and is shown in Figure~\ref{taus}(a) for the spatial coordinate of $z = 1.24$\,mm at 10\,kHz frequency of the pulse repetition. The temporal stability of the $\tau_{\mathrm{eff,1}}^{2p_i}(t)$ during the electric field rise time is apparent.

% \begin{figure}[h!]
% \begin{center}
% a)
% \includegraphics[clip = true,width=0.45\columnwidth]{f10kHz-ns-124-tau1.png}
% b)
% \includegraphics[clip = true,width=0.45\columnwidth]{f10kHz-ns-124-tau2.png}
% c)
% \includegraphics[clip = true,width=0.45\columnwidth]{f10kHz tau2 determination 124mm.png}
% d)
% \includegraphics[clip = true,width=0.45\columnwidth]{Tau3_124-2p2-2p5.png}
% e)
% \includegraphics[clip = true,width=0.45\columnwidth]{f10kHz-ns-124-tau3.png}
% \end{center}
% \caption{
% taus for theory of experimental ns pulse 10kHz 1.24mm
% }
% \label{taus}
% \end{figure}

Second, we include all processes $S_{2p_i}(z, t)$ contributing to the population and depopulation of the 2p$_{i}$ state density, except the direct electron impact driven process given by the rate $k_{g,2p_i}(E(z,t)/N)$. The latter 
%which 
is given separately in the balance equation. %~\eref{tau1eq}. 
This should be the best option, as it 
%is describing 
describes 
the state density with the accuracy of %the 
equation \eref{population}. 
%Yet for 
For 
experimental uses, such parameter is basically  inaccessible due to its rapid temporal variation. 
We denote it as type 2 effective lifetime $\tau_{\mathrm{eff,2}}^{2p_i}(z,t)$ given 
by 
%as follows:

\begin{equation}
\label{tau2eq}
\tau_{\mathrm{eff,2}}^{2p_i}(z,t) = \frac{1}{S_{2p_i}(z, t)-k_{g,2p_i}(E(z,t)/N)\cdot n_e(z,t)\cdot n_g}.
\end{equation}

\noindent
The results for the same conditions as for the type 1 lifetime are shown in Figure~\ref{taus}(b). The type 2 lifetimes are negative in most of their temporal range because the excitation rates are larger than the other loss processes in $S_{2p_i}(z,t)$ and result in sharp peaks where direct excitation to the 2p$_i$ state dominates the source term $S_{2p_i}$ (and the definition \eref{tau2eq} becomes problematic due to division by zero). Due to these complex properties, the $\tau_{\mathrm{eff,2}}^{2p_i}(z,t)$ 
%will not be 
is not 
used in the presented evaluations and its more detailed analysis is left for future work.

%The oscillation of their values for all 2p states is apparent. For their application into the equation~\eref{population2} we have determined the average value of $\tau_{\mathrm{eff,2}}^{2p_x}(z,t)$ averaging over the enhanced electric field of the streamer head, i.e.\ typically over 20\,ns, see Figure~\ref{taus}c). The interval used for averaging is denoted by two vertical black lines (before 10 and after 30\,ns).

\begin{figure}[tb]
\begin{center}

\includegraphics[clip = true,width=0.95\columnwidth]{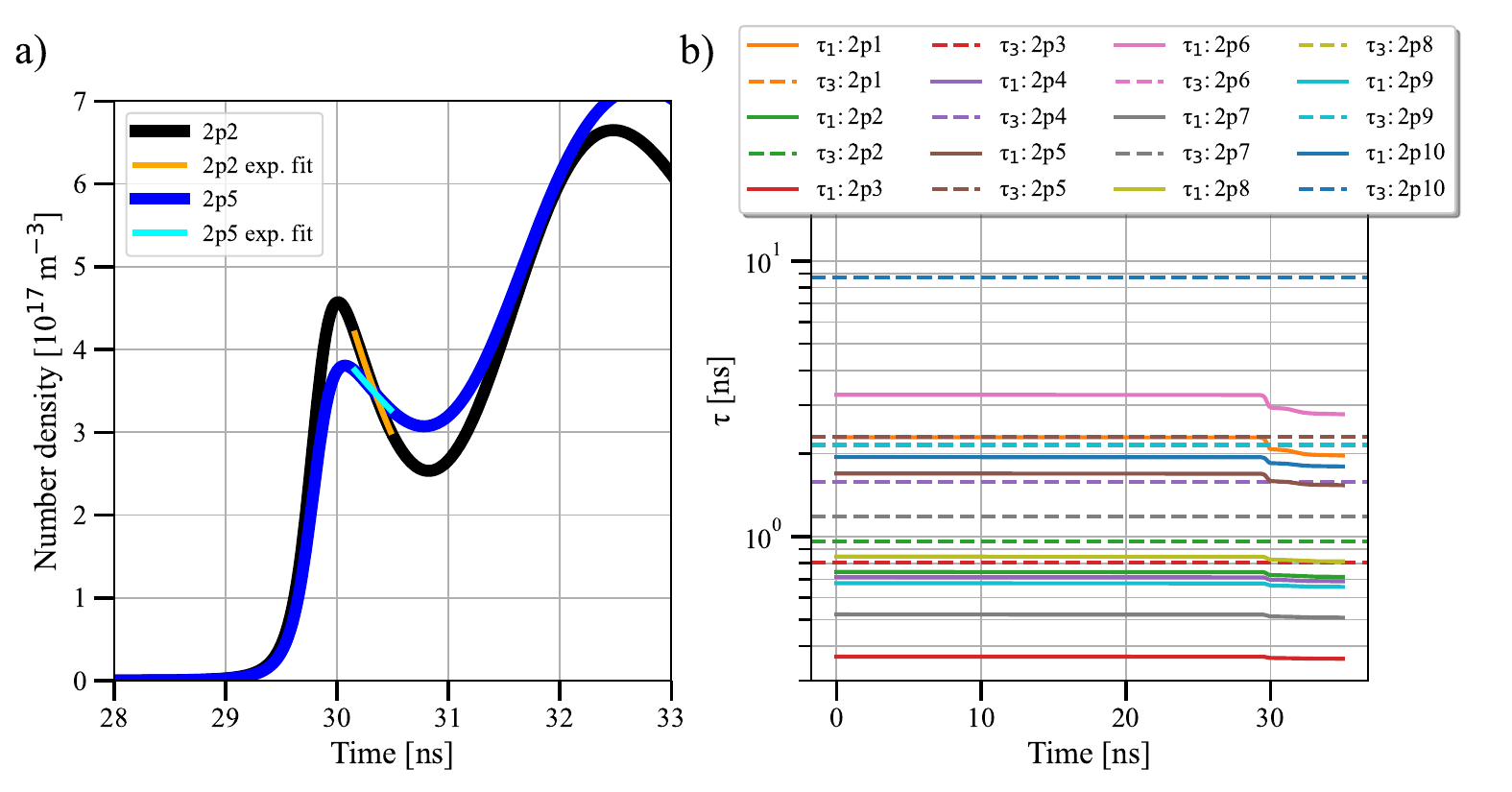}

\end{center}
\caption{
Illustration of exponential fitting procedure used to determine $\tau_{\mathrm{eff,3}}^{2p_i}(z,t)$ (a) and comparison of $\tau_{\mathrm{eff,1}}^{2p_i}(z,t)$ and $\tau_{\mathrm{eff,3}}^{2p_i}(z,t)$ (b). The figures represent the number densities and effective lifetimes for a discharge frequency of 10\,kHz and spatial coordinate of $z = 1.24$\,mm.
Precise values of $\tau_{\mathrm{eff,1}}^{2p_i}(z,t)$ and $\tau_{\mathrm{eff,3}}^{2p_i}(z,t)$ are summarized in Table\,\ref{tab:tau1_tau3_lifetimes}. % 
%\DLR{Remark DL to (a): There are too many points, which are full as well. The experimental fit is very difficult on screen and is not recognizable in printed form, where black dots overright everything. Less open circles would be better. } 
}
\label{taus_tau1_vs_tau3}
\end{figure}

In a third approach, we quantify the effective lifetimes of type 3,  $\tau_{\mathrm{eff,3}}^{2p_i}(z,t)$, from the exponential decay of the 2p state densities after the streamer head passage. 
This is the only experimentally accessible effective lifetime. It can be measured with a detector of a sufficient, typically sub-nanosecond,  temporal resolution, i.e.\ streak camera, fast photomultiplier, fast gated intensified CCD camera or a time-correlated single photon counting technique. 
Effective lifetimes and quenching coefficients were experimentally determined by this method already previously, see for example \cite{pancheshnyi1998,gans2003,kozlov2005}. 
This evaluation procedure is illustrated together with its results in Figure~\ref{taus_tau1_vs_tau3} 
%(a) and (b) 
for the investigated discharge and compared to the effective lifetime of type 1 according to \ref{tau1eq}. 
The determined values of $\tau_{\mathrm{eff,1}}^{2p_i}(z,t)$ and $\tau_{\mathrm{eff,3}}^{2p_i}(z,t)$ are also given in Table\,\ref{tab:tau1_tau3_lifetimes}. It should be noted that in some cases the decay of the 2p state densities after the streamer head was overlapped with the later discharge phase, making the determination of $\tau_{\mathrm{eff,3}}^{2p_i}(z,t)$ impossible. These values are missing in 
Table\,\ref{tab:tau1_tau3_lifetimes} and Figure~\ref{taus_tau1_vs_tau3}. 
The overlap problem is also the reason for the failure of this effective lifetime determination in the real experiments, see further in the text.

\begin{table*}[htb]
\center
\caption{Effective lifetimes $\tau_{\mathrm{eff,1}}^{2p_i}(z,t)$ and $\tau_{\mathrm{eff,3}}^{2p_i}(z,t)$ for a discharge frequency of 10\,kHz. The values for spatial coordinate 1.24\,mm are also depicted in Figure~\ref{taus_tau1_vs_tau3}(b). The values that could not be determined because of the streamer decay overlapping with subsequent discharge phases are not presented, including the entire spatial position of 1.48\,mm. The spatial positions of the used $\tau_{\mathrm{eff,1}}^{2p_i}(z,t)$ are not specified as the values for individual positions are identical before the discharge.}
\label{tab:tau1_tau3_lifetimes}

\begin{tabular}{lcccccccccc}\br
Lifetime [ns] & 2p$_1$ & 2p$_2$ & 2p$_3$ & 2p$_4$ & 2p$_5$ & 2p$_6$ & 2p$_7$ & 2p$_8$ & 2p$_9$ & 2p$_{10}$ \\\mr
$\tau_1$ & 2.29   & 0.74   & 0.37   & 0.71   & 1.69   & 3.27   & 0.52   & 0.84   & 0.67   & 1.94      \\
$\tau_3$ (1\,mm) & -      & 1.95   & 1.37   & 2.7    & 9.8    & -      & 2.97   & 8.67   & 8.16   & -         \\
$\tau_3$ (1.24\,mm) &   -      & 0.96   & 0.80   & 1.58   & 2.31   & -      & 1.18   & 2.14   & 2.15   & 8.72      \\
\hline
\end{tabular}
\end{table*}

% \begin{figure}[h!]
% \begin{center}
% a)
% \includegraphics[clip = true,width=0.45\columnwidth]{Int_vs_der_2p5-124-teory.png}
% b)
% \includegraphics[clip = true,width=0.45\columnwidth]{E_field_2p5-124-teory.png}
% c)
% \includegraphics[clip = true,width=0.45\columnwidth]{E_field_f10kHz-124-10kHz-point-coef.png}
% d)
% \includegraphics[clip = true,width=0.45\columnwidth]{2D_overview_logscale.png}
% \end{center}
% \caption{vysledky ns.p. theory tau1 na 1.24 a) b) point a c) ziskani koeficientu d) prehledovy obrazek z 1D modelu,  10kHz  
% analysis1}
% \label{analysis1}
% \end{figure}

\begin{figure}%[h!]
\begin{center}

\includegraphics[clip = true,width=0.95\columnwidth]{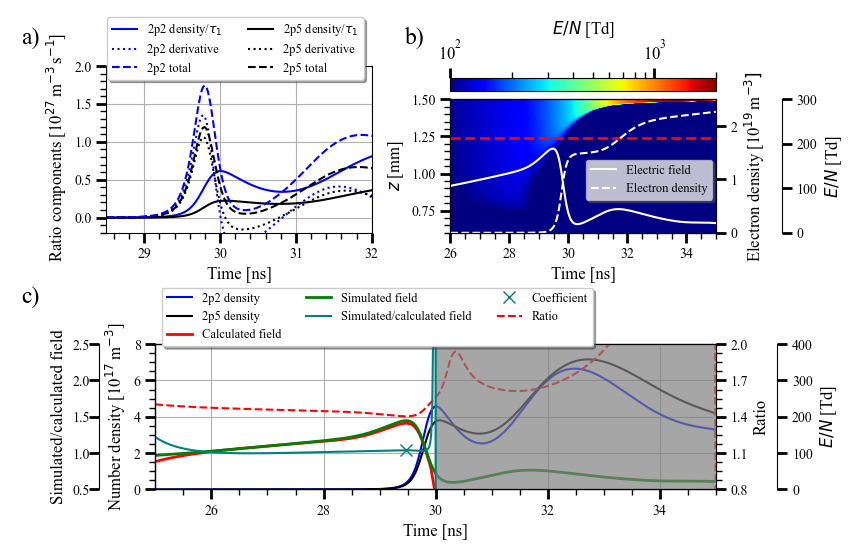}

\end{center}
\caption{
%\textcolor{red}{This caption is much too long and confusing. Please try to explain things throughout the discussion in the text.}
Determination of $E/N$ and $E/N$ multiplication coefficient from the ratio of 2p$_2$ to 2p$_5$ states from the 1D model. 
Figure (a) shows the temporal development of the components used for determination of $E/N$.
%: number densities divided by the effective lifetime $\tau_{\mathrm{eff,1}}^{2p_i}(z,t)$, derivatives of the number densities and their sum, i.e.\ the whole numerator or denominator from equation \eref{ratioeq}, denoted as total. 
Figure (b) displays the spatiotemporal behaviour of $E/N$ from the 1D model for 10\,kHz together with the profile of the $E/N$ and electron number density at 1.24\,mm.
%Note that only a part of the gas gap is presented from 0.6 to 1.5\,mm to highlight the area of interest. 
The spatial position 1.24\,mm is depicted as the red dashed line and all of the displayed data in figure (a) and (c) corresponds to this position. In figure (c) the ratio of simulated to calculated field is denoted 
%\DL{by the teal cross,} 
as a teal waveform, 
where the ratio of their maxima ($E/N$ multiplication coefficient) is denoted as a \ja{teal cross}. The red dashed line represents the ratio of total components in figure (a), i.e.\ the left side of equation \eref{ratioeq}. \ja{The shaded area in (c) corresponds to the discharge phase where direct electron impact excitation is not dominant (after the streamer head passed the spatial point) and therefore the presented method is not applicable, i.e. the range is out of interest for further analysis.}}
\label{analysis1}
\end{figure}

Having direct access to the defined effective lifetimes, we have performed a semi-automated evaluation of %the 
equation~\eref{ratioeq} 
using  
these lifetimes and used the 2p states densities obtained from the 1D  simulations (point data) as well. 
The 2p state densities were available for different frequencies and coordinates in the gap for the nanosecond pulsed barrier discharge. 
An example of such evaluation for the 10\,kHz repetitive discharge and the spatial coordinate of $z$ = 1.24\,mm is shown in Figure~\ref{analysis1}, where the first type of effective lifetime $\tau_{\mathrm{eff,1}}^{2p_i}(z,t)$ is used to evaluate the ratio of state densities 2p$_2$/2p$_5$.

In Figure~\ref{analysis1}(a), the temporal developments of the densities of radiative states 2p$_2$ and 2p$_5$ (divided by the respective effective lifetimes $\tau_{\mathrm{eff,1}}^{2p_2}(z,t)$ and $\tau_{\mathrm{eff,1}}^{2p_5}(z,t)$) are shown, together with their derivatives and sums of these two components (total), for the coordinate of $z$ = 1.24\,mm. 
The coordinate is highlighted 
by the red dashed line in Figure~\ref{analysis1}(b) for better orientation. 
As also shown in \cite{goldberg2022}, the density derivatives can be a dominant part of the left-hand side of equation \eref{ratioeq}.  As it is obvious from Figure~\ref{analysis1}(a), they completely dominate at the very beginning. It is obvious, as we have a rapidly changing electric field in the passing fast streamer induced by the nanosecond pulse of applied voltage. 

%\textcolor{red}{Introduction and discussion of Figure \ref{analysis1}(b) are missing. }
% The $E/N$ from 1D model is depicted in Figure~\ref{analysis1}(b) together with the temporal development of $E/N$ and $n_e$ for selected position of $z=1.24\,$mm.

In Figure~\ref{analysis1}(c), the original simulated electric field from the fluid model simulation is shown (green) together with the electric field calculated from the equation~\eref{ratioeq} (red), using the selected effective lifetimes and the excitation rate coefficients as described earlier in the text. 
The ratio of simulated to calculated $E/N$ peak values is evaluated as so-called $E/N$ multiplication coefficient, which further serves for quantification of the accuracy using equation~\eref{ratioeq} with given state densities and effective lifetimes for $E/N$ determination. 
Obviously, the closer the multiplication coefficient is to unity, the better equation~\eref{ratioeq} approximates the complex generation and loss processes for the respective 2p state in the discharge under given conditions. 
\ja{As it was previously discussed, the presented intensity ratio method requires direct electron impact excitation from the ground state to be a dominant population process for the considered 2p states (all other processes are considered in an effective lifetime, see equation~\ref{population}). 
However, this assumption is not valid after the streamer head has passed (approx. at 30\,ns) because then stepwise excitation and quenching processes start to contribute significantly to the population of all 2p states. In this temporal region, marked as a shaded region in Fig.\,\ref{analysis1}(c), the calculated $E/N$ represents an artefact and should not be considered for further analysis. More discussion regarding this artefact and its identification can be found for example in~\cite{goldberg2022,kozlov2001,hoder2015}.}
% The peak values of both electric field waveforms are highlighted by a circular point and the ratio simulated/determined is evaluated as so called $E/N$ multiplication coefficient, which further serves for quantification of the accuracy using equation~\eref{ratioeq} with given states and lifetimes for $E/N$ determination. Obviously, the closer the multiplication coefficient is to unity, the better the simple  equation~\eref{ratioeq} represents the complex plasma processes in the discharge. 

% \begin{figure}[h!]
% \begin{center}
% a)
% \includegraphics[clip = true,width=0.95\columnwidth]{coef-ns-2p2.png}
% b)
% \includegraphics[clip = true,width=0.95\columnwidth]{coef-ns-2p3.png}
% c)
% \includegraphics[clip = true,width=0.95\columnwidth]{coef-ns-2p4.png}
% \end{center}
% \caption{
% coefficients tau1 ns pulsed, 1 and 1.24 and 1.48mm positions of the columns}
% \label{coef1}
% \end{figure}

\begin{figure}%[h]
\begin{center}

\includegraphics[clip = true,width=0.95\columnwidth]{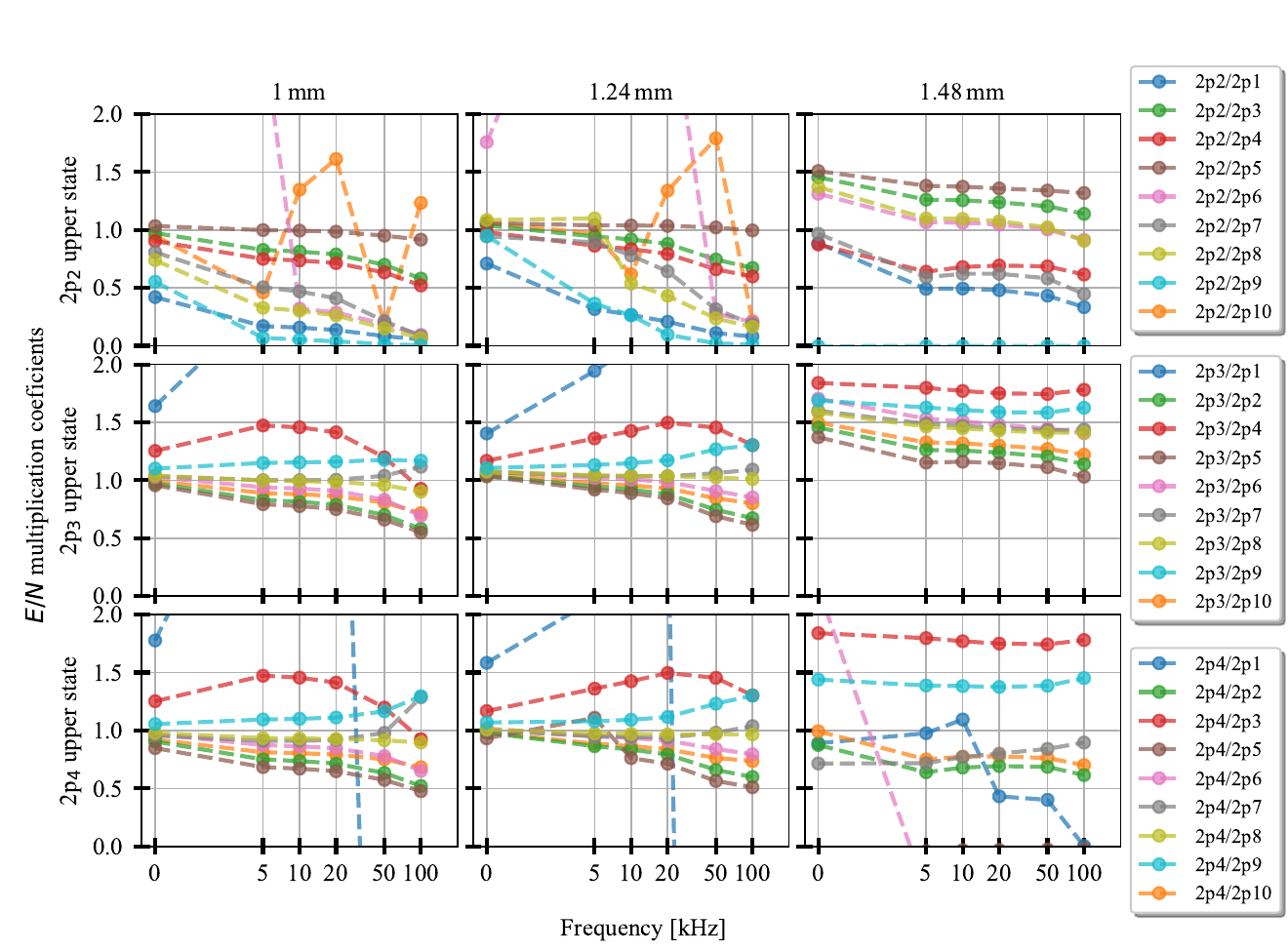}

\end{center}
\caption{
$E/N$ multiplication coefficients with the application of $\tau_{\mathrm{eff,1}}^{2p_i}(z,t)$. The coefficients are displayed for all combinations of 2p$_{1-10}$ states with 2p$_2$, 2p$_3$, and 2p$_4$ upper states, they are displayed as individual rows. The columns represent spatial positions 1, 1.24, and 1.48\,mm. Calculations are conducted 
\ja{based on results of} 
%for 
1D simulations for frequencies of 5, 10, 20, 50 and 100\,kHz and also for the first discharge depicted as 0\,kHz. The frequencies are displayed in logarithmic scale with the interval from 0 to 5\,kHz being linear. }
\label{coef1}
\end{figure}

We have determined the multiplication coefficient for $E/N$ during the streamer propagation phase in the nanosecond barrier discharge using a semi-automated procedure at different frequencies and for three spatial coordinates.
The results are shown in Figure~\ref{coef1}.  
Here we want to find the best suitable 2p states ratio by a ``computational force" of simulating the discharge and evaluating the $E/N$ using the equation \eref{ratioeq} for different conditions and scenarios. 
This procedure 
\ja{has} 
%had 
the following reasons. The different frequencies represent the different preionization and initial metastable state densities in the gas gap where the streamer propagates, as it was quantified earlier in subsection~\ref{sec:different_frequencies}. 
The interval of their values is given by at least $2\times 10^{9}$ (initial conditions for the first discharge) and at most $4.73\times 10^{17}\,$m$^{-3}$ (initial conditions for the 100\,kHz discharge) for electron density and $10^{9}\,$m$^{-3}$ at least and $10^{15}\,$m$^{-3}$ at most for the metastable states densities in the evaluated coordinates. 
The different coordinates of the $E/N$ evaluation for all these frequencies then give the sensitivity of the searched 2p states ratio to the electric field of different amplitudes. 
The evaluated interval of the amplitudes is from 100 to 2200\,Td, approximately.

% The automated procedure had a following structure:
% ...
The semi-automated procedure consists of following phases: The first phase of the semi-automated procedure 
\ja{is} 
%was 
the determination of the $E/N$ based on equation~\eref{ratioeq} by the use of $\tau_{\mathrm{eff}}^{2p_i}(z,t)$ (type 1 and 3), see Figure~\ref{taus_tau1_vs_tau3}, and the use of number densities of respective 2p states. 
The individual components of this evaluation can be seen in Figure~\ref{analysis1} with the resulting ratio of the left-hand side of equation~\eref{ratioeq} in Figure~\ref{analysis1}(c) depicted as a red dashed line. 
The determination of $E/N$ from the density ratios is based on the rate coefficients shown in Figure~\ref{ratios}. 

The second phase of the procedure is concerned with the evaluation of the $E/N$ multiplication coefficient, i.e.\ the ratio of peak values of the simulated and the calculated electric field. 
The procedure is however not as straightforward for the calculated fields as their temporal profile often displays local variations that tend to disrupt the process. 
One of the main reasons for these deviations is the low sensitivity of ratios of rate coefficients to $E/N$. 
This often leads to ambiguously defined $E/N$ that fluctuate between multiple values. % as the calculated ratio corresponds to the multiple values of the ratio of rate coefficients. 
This is mostly countered by using extrapolated values of the ratio of rate coefficients for a selected range of $E/N$, typically 100 to 1000\,Td (up to 2000\,Td), see Figure~\ref{ratios}. 
This is important information also for the possible experimental use of the intensity ratio. Above some $E/N$ values the function of rate coefficients ratio 
\ja{changes} 
%is changing 
the polarity of its derivative and is not ambiguous with respect to the $E/N$. 
%that around 100 and 100\,Td is often an inflection point.
Another aspect to consider is the complexity of the temporal development of the calculated $E/N$.
In the temporal range with low number densities of 2p states, the calculated $E/N$ is often not very precise and significant fluctuations take place. 
Similar deviations are also present after the maximum of the streamer head where the 
\ja{CRM of the line ratio method using equation~\eref{ratioeq}}
%utilized kinetic model 
%\DLR{Remark DL: is that what you mean by the former wording utilized kinetic model?} 
is no longer valid. Such problems are also present in the evaluation of experimental data, see e.g.\ \cite{goldberg2022}.

To suppress any persistent deviations of calculated $E/N$, only the very close proximity of the maximum of the simulated field is used to find the maximum of the calculated field to compare with. The calculated $E/N$ in this temporal range is Gaussian smoothed and its time derivative is tracked. 
The maximum is marked when the derivative is equal to zero. 
%The time derivatives are also used to track any possible fluctuations or complex temporal development of calculated $E/N$ and to identify their "correct" $E/N$ value based on the shape and similarity of the calculated field to the simulated field.
It should be noted that the evaluation parameters (especially the temporal ranges tracked by time derivative) were set manually for every spatial position to increase the precision of the procedure.
Even after all the corrections described above, the influence of the low sensitivity of ratios of rate coefficients on $E/N$ and complex temporal development of the calculated $E/N$ was still not completely suppressed. However, after a detailed investigation, it was concluded that only a few ambiguously defined $E/N$ are present and mostly for the combinations among states 2p$_{1, 6-10}$.  
% The resulting series of $E/N$ multiplication coefficients were not error-free but after detailed investigation, it was concluded that only a few errors took place mostly for the combinations among states 2p$_{1, 6-10}$. 
These problematic coefficients usually manifest themselves by their oscillatory nature as functions of frequency (see e.g.\ the coefficients for 2p$_2$/2p$_{10}$ for position 1.24\,mm in 
\ja{the centred part of the first row of} 
Figure~\ref{coef1}). The oscillatory behaviour also takes place for the ratio of states whose calculated $E/N$ is significantly higher or lower than that of the simulated field. Those are typically the states in combination with 2p$_1$.

\begin{figure}%[h]
\begin{center}

\includegraphics[clip = true,width=0.7\columnwidth]{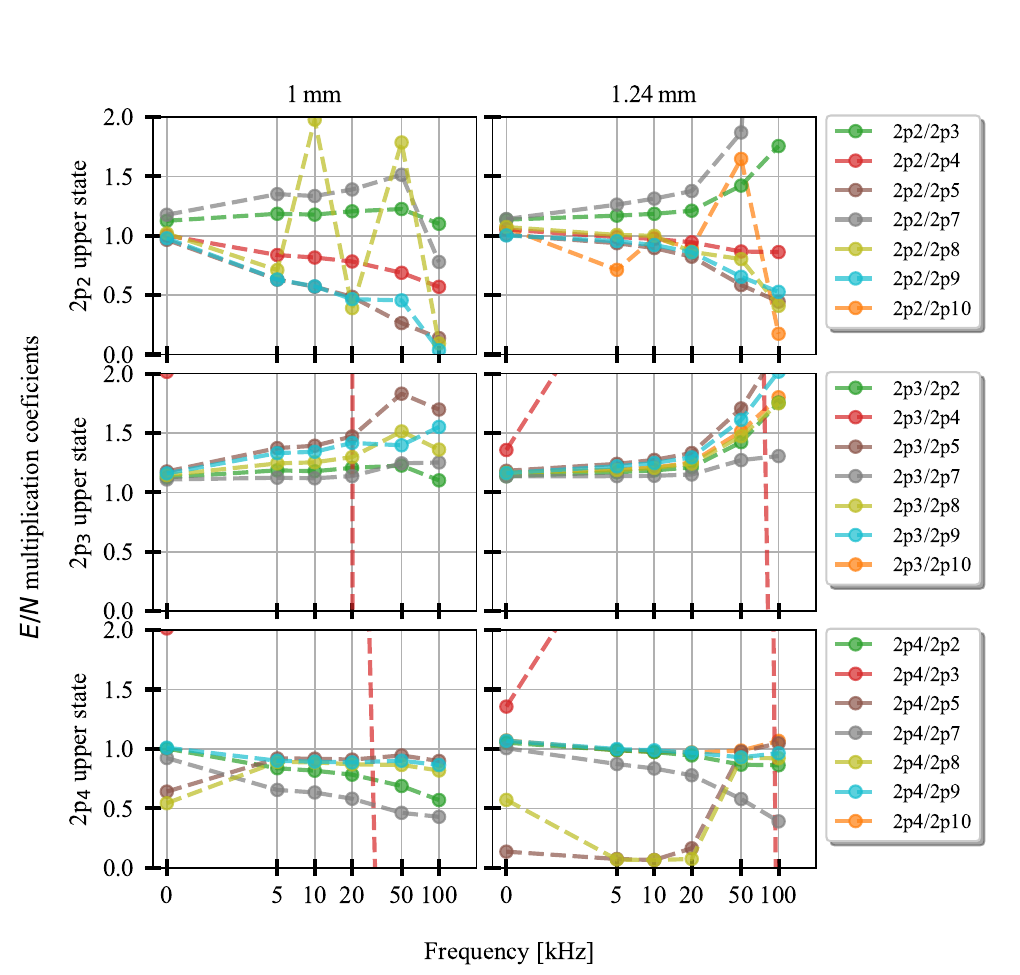}

\end{center}
\caption{
$E/N$ multiplication coefficients with the application of $\tau_{\mathrm{eff,3}}^{2p_i}(z,t)$. The coefficients are displayed for all combinations of 2p$_{1-10}$ states with 2p$_2$, 2p$_3$, and 2p$_4$ upper states, they are displayed as individual rows. The columns represent spatial positions 1 and 1.24\,mm. Calculations are conducted for 1D model at frequencies 5, 10, 20, 50 and 100\,kHz and also for a first discharge depicted as 0\,kHz. Note that coefficients for position 1.48\,mm, ratio of states including 2p$_1$, 2p$_6$ and partly 2p$_{10}$ are not displayed as their respective $\tau_{\mathrm{eff,3}}^{2p_i}(z,t)$ 
\ja{could not be} 
%have not been 
determined, see table\,\ref{tab:tau1_tau3_lifetimes}. The frequencies are displayed on a logarithmic scale with the interval from 0 to 5\,kHz being linear.}
\label{coef3}
\end{figure}

By careful analysis of the results, we have found that using effective lifetimes of the first type, $\tau_{\mathrm{eff,1}}^{2p_i}(z,t)$, the coefficient is closer to unity for most frequencies and positions dominantly for the ratios including the states 2p$_2$, 2p$_3$ and 2p$_4$. It was already previously estimated that these states should be dominantly populated by the direct electron impact excitation from the ground state using optical emission spectroscopy for the same nanosecond barrier discharge in argon, see \cite{kusyn2023}.
Moreover, the ratio of 2p$_2$ to 2p$_5$ was found to be relatively stable for all three positions and over the investigated interval of pulse repetition frequencies. Ratios using the densities of 2p$_8$ state in combination with 2p$_3$ and 2p$_4$ resulted also 
\ja{in an} 
%to 
$E/N$ multiplication coefficient close to unity. 
%\DLR{Comment DL: I propose to add the following sentence added in red, if it is correct. 
\ja{Thus, they might also become interesting for the use of the intensity  ratio method and will be analysed in more detail in future studies.}

% \begin{figure}[h!]
% \begin{center}
% a)
% \includegraphics[clip = true,width=0.28\columnwidth]{E_field_2p2 tau2 1mm.png}
% b)
% \includegraphics[clip = true,width=0.28\columnwidth]{E_field_2p2 tau2 124mm.png}
% c)
% \includegraphics[clip = true,width=0.28\columnwidth]{E_field_2p2 tau2 148mm.png}
% \end{center}
% \caption{
% coefficients tau2 ns pulsed, 1 and 1.24 and 148mm positions of the columns}
% \label{coef2}
% \end{figure}

% \begin{figure}[h!]
% \begin{center}
% a)
% \includegraphics[clip = true,width=0.28\columnwidth]{E_field_2p2 tau3 1mm 10kHz.png}
% b)
% \includegraphics[clip = true,width=0.28\columnwidth]{E_field_2p2 tau3 124mm 10kHz.png}
% c)
% \includegraphics[clip = true,width=0.28\columnwidth]{E_field_2p2 tau3 148mm 10kHz.png}
% \end{center}
% \caption{
% coefficients tau3 ns pulsed, 1 and 1.24 and 148mm positions of the columns}
% \label{coef3}
% \end{figure}

Evaluating the $E/N$ multiplication coefficients also for the third type of the effective lifetimes, i.e.\ $\tau_{\mathrm{eff,3}}^{2p_i}(z,t)$, we came to very similar results, see Figure~\ref{coef3}. The figure shows the coefficients for all combinations of states with the upper states 2p$_2$, 2p$_3$ and 2p$_4$ for positions 1 and 1.24\,mm. The ratios containing the states 2p$_1$, 2p$_6$ and partly 2p$_{10}$ are excluded from the evaluation, as it was not possible to determine positive values of their effective lifetimes, as shown in the table\,\ref{tab:tau1_tau3_lifetimes}.
% The figure is missing the coefficients for position 1.48\,mm and ratios of states including 2p$_1$, 2p$_6$, and partly 2p$_{10}$ as it was not possible to determine positive values of the effective lifetimes as is shown in table\,\ref{tab:tau1_tau3_lifetimes}. 
However, based on the relatively good match of $E/N$ multiplication coefficients for the remaining theoretical lifetimes of the type 3, we can state that also this approach can deliver reasonable $E/N$ values. The ratio of 2p$_4$ to 2p$_9$ results in a good $E/N$ computation over all frequencies and both positions.

% Dominantly the coefficients including the 2p2 state are shown as they were the closest to unity and relative stable over the investigated range of positions and frequencies. The ratio of 2p2 and 2p4 states was found well stable additionally. 

\subsection{Comparison with experimental data and effect of spatial integration of radiative state densities}\label{sec:Comparison}

% The two robust ratios of 2p2/2p5 and 2p2/2p4 were used for evaluation of the experimental data presented in \cite{kusyn2023}. 
% field averaged and densities integrated
% for 1D and 2D
% !!!
% tau2 not good for experiment
% at least 2p2/2p5 ... better to try 2p2/2p4 
% on axis max field is 589, integrated maximal field is 557td

\begin{figure}%[h!]
\begin{center}

\includegraphics[clip = true,width=.6\columnwidth]{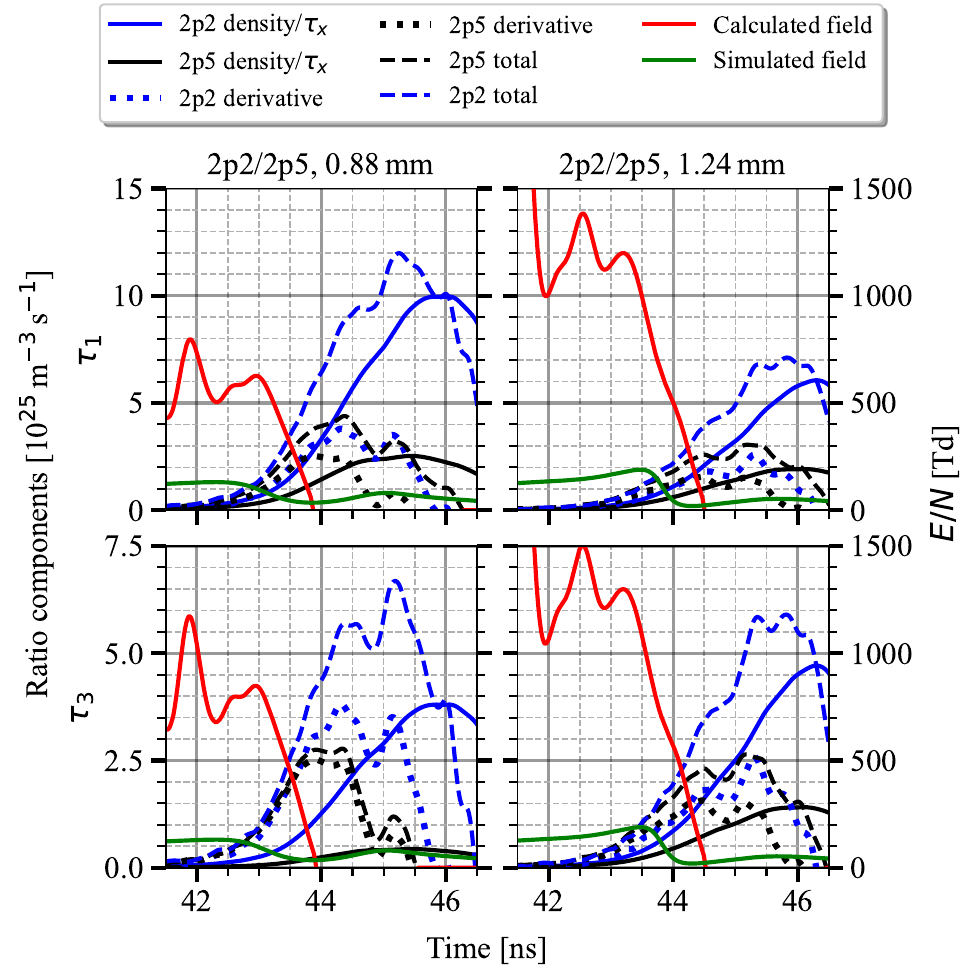}

\end{center}
\caption{Smoothed components necessary for intensity ratio technique, see equation~\eref{ratioeq}, with resulting $E/N$. The figure shows the results from an evaluation of ratio 2p$_2$/2p$_5$ %and 2p$_2$/2p$_4$ 
at two spatial positions 0.88 and 1.24\,mm. Calculations were conducted on measured data discussed in detail in\,\cite{kusyn2023} with application of $\tau_{\mathrm{eff,1}}^{2p_i}(z,t)$ and $\tau_{\mathrm{eff,3}}^{2p_i}(z,t)$. The rows represent calculations with respective effective lifetimes, while the columns represent individual spatial coordinates. The simulated (green) $E/N$ is not temporally correlated with the calculated field and is only used for a comparison of fields maxima, not temporal position.
%\textcolor{blue}{I have taken our ratio 2p2/2p4, but the figure including it is in files as Figure10.pdf.}
}
\label{field_comparison_measured_data}
\end{figure}

The relatively robust ratio of 2p$_2$ to 2p$_5$ was used for evaluation of the experimental data presented in\,\cite{kusyn2023}. 
The same procedure for the determination of $E/N$, as shown in Figure~\ref{analysis1}, is conducted on measured number densities using the equation~\eref{ratioeq} and both proposed lifetimes. 
The resulting $E/N$ compared with the $E/N$ obtained from the 1D simulation is shown in Figure~\ref{field_comparison_measured_data} together with the individual components on left hand side of equation~\eref{ratioeq}.
This figure 
\ja{also compares} 
%is also comparing 
the influence of theoretically obtained effective lifetimes of type 1 and 3 on computed $E/N$, with fairly similar results of difference $\sim 200\,$Td for 0.88\,mm (experimental position 0.88\,mm was compared with the simulated position of 1\,mm) and $\sim 100\,$Td for 1.24\,mm.
It is important to mention that the relevant value of $E/N$ in the streamer head is located in the proximity of 43\,ns. 
Notice that similarly as in the case of simulated data, the 
\ja{time}  
derivatives of the densities of 2p states are an important contribution to the total quantity (i.e.\ to numerator or denominator of the equation~\eref{ratioeq}). 
Probably due to the limited temporal resolution or lower signal-to-noise ratio in 
\ja{the} 
%those  
measurements, the derivatives are not as dominant as in the case of simulated data.
As the values of 2p states' number densities are in that moments relatively low, they are prone to fluctuations and subsequently so are the derivatives.
Nevertheless, despite the fluctuations of the derivatives and the low sensitivity of $E/N$ on the ratio of rate coefficients, the proposed ratio 2p$_2$/2p$_5$ is in qualitative agreement with the simulated field for the two positions. 
\ja{Figure~\ref{field_comparison_measured_data} also indicates that the $E/N$ increases for coordinates closer to the cathode, as expected, showing that even in experimental conditions the determination of $E/N$ by the intensity ratio method is possible, i.e. sensitive to $E/N$ variations.} 
This suggests that the ratio 2p$_2$/2p$_5$ may be one of the best options found for $E/N$ determination by a presented method, even if the absolute values are unphysically high.

% \begin{center}
% a)
% \includegraphics[clip = true,width=0.45\columnwidth]{Int_vs_der_2p5-088.png}
% b)
% \includegraphics[clip = true,width=0.45\columnwidth]{E_field_2p5-088.png}
% c)
% \includegraphics[clip = true,width=0.45\columnwidth]{E_field_2p5-044.png}
% d)
% \includegraphics[clip = true,width=0.45\columnwidth]{E_field_2p5-124.png}
% \end{center}
% \caption{vysledky ns.p. experimental 0.88 dvakrat pak c) 0.44 a d) je 1.24
% analysis   --- tau1}
% \label{analysis}
% \end{figure}

An important aspect that should be considered is that the optical emission spectroscopy measurement is a line-of-sight integration detection. As mentioned previously, due to its microscopic diameter, the discharge filament is projected perpendicularly onto the monochromator slit. As a result, the measured intensities/densities are axially and radially integrated. To resolve this issue, the data obtained by using 1D and 2D model calculations are analysed in the further part of the manuscript.

\begin{figure}%[h!]
\begin{center}
\includegraphics[clip = true,width=.7\columnwidth]{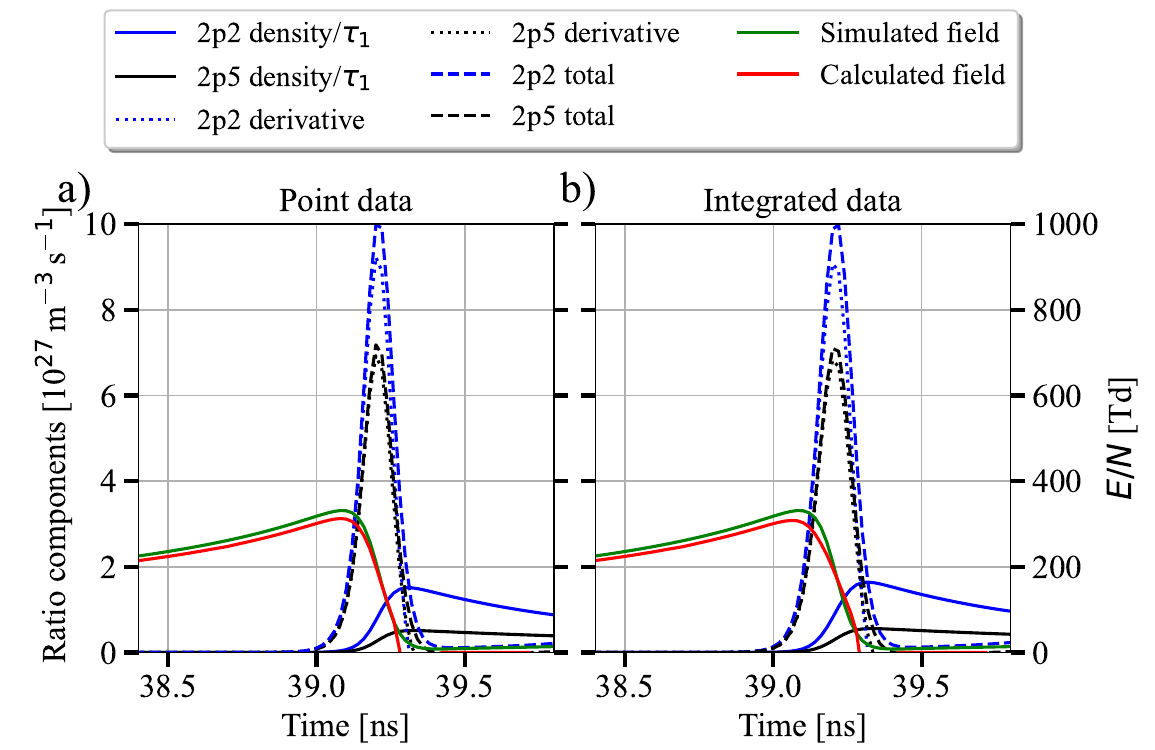}
\end{center}
\caption{Comparison of ratio components of point data and their integrated values over an axial interval of 30$\,\mu $m for first discharge and position of 1.24\,mm. Data are from a 1D simulation. The integrated data are in m$^{-2}$s$^{-1}$ and are scaled by the ratio of the maxima of 2p$_2$ totals for illustrative purposes. The simulated field (green curves) corresponds to the $E/N$ of point data for both parts of the graph.}
\label{integrated_vs_onaxis_1D}
\end{figure}

\begin{figure}%[h!]
\begin{center}
\includegraphics[clip = true,width=.7\columnwidth]{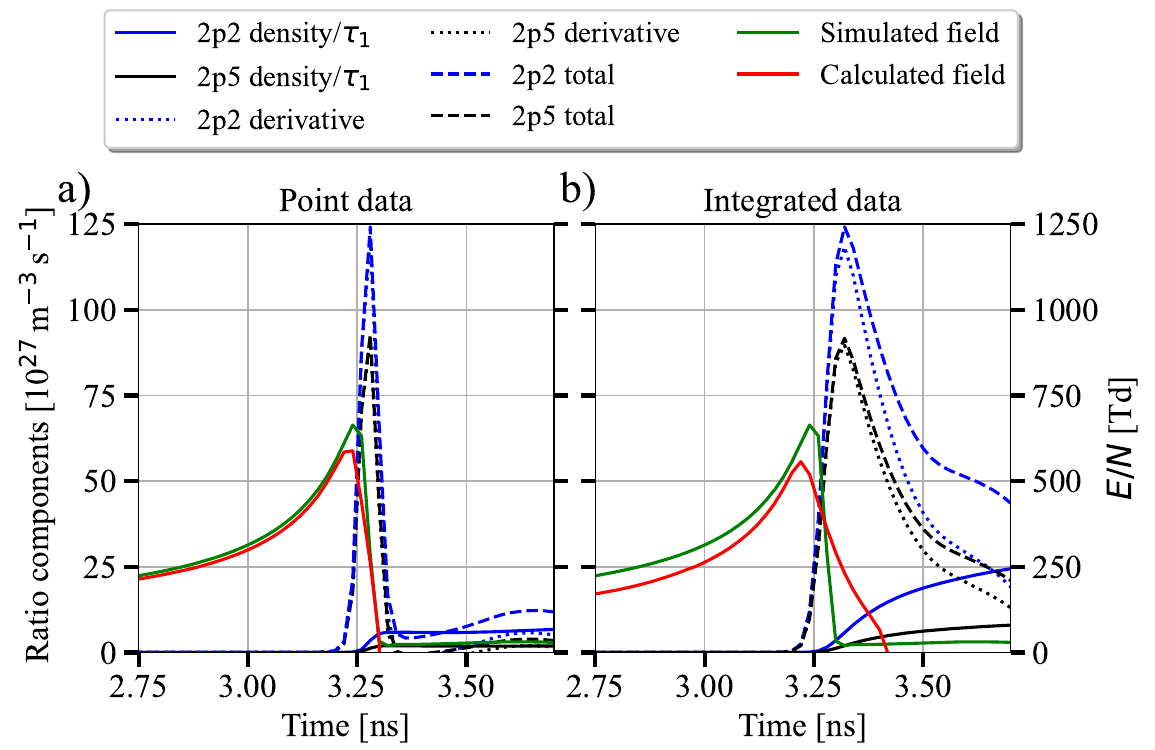}
\end{center}
\caption{Comparison of ratio components of point data and their integrated values over a volume interval of 30$\,\mu $m for first discharge and position of 1.24\,mm. Data are from a 2D simulation. The integrated data are in s$^{-1}$ and are scaled by the ratio of the maxima of 2p$_2$ totals for illustrative purposes. The simulated field (green curves) corresponds to the $E/N$ of point data for both parts of the graph.}
\label{integrated_vs_onaxis_2D}
\end{figure}

The so far discussed point data 
(2\,$\mu$m cell on axis) of number densities are presented and compared with $E/N$ computations from axially integrated (for 1D model, see Figure~\ref{integrated_vs_onaxis_1D}) and with volume integrated number densities (see Figure~\ref{integrated_vs_onaxis_2D} for 2D model) for the first discharge at the position of 1.24\,mm. The volume integration is given by equation \eref{tau}. 
Both figures depict the individual ratio components, their calculated $E/N$ (with an effective lifetime of type 1), and the simulated field. 
%of axial data for both the axial and integrated data of 2p states densities. 

In Figure~\ref{integrated_vs_onaxis_1D}, the comparison between point data and axially integrated data over 30\,$\mu$m is shown. It shows that if the 1D simulated number densities are integrated in the axial dimension, the respective calculated $E/N$ is almost identical to the simulated $E/N$ with deviation in units of Td for 1 and 1.24\,mm for the selected ratio 2p$_2$/2p$_5$. 
The values of the calculated field for point data in Figure~\ref{integrated_vs_onaxis_1D} is 313\,Td and the value for integrated data is 308\,Td; in comparison the value of the simulated field is 331\,Td. 
Therefore, the deviation of the calculated fields from the simulated field is 18.6\,Td for point data densities and 23.5\,Td for integrated data densities.
The significant difference occurs only for the case of 1.48\,mm, presumably because it is only 0.02\,mm from the electrode and the integration includes an area corresponding to the cathode spot, i.e.\ an area with a high spatiotemporal gradient of $E/N$.

In Figure~\ref{integrated_vs_onaxis_2D} the comparison of point data and volume integrated data is shown. 
We can see that the match of the calculated and simulated field is not as good for integrated data as for point data. 
The integration of the densities using equation \eref{tau} results in computed $E/N$ with lower amplitude and broader waveform. 
However, the amplitude deviation is still only 16\,\%. 
Also, the influence of time derivatives is not as influential in the case of integrated data as in the case of point data evaluation and is similar to experimental data evaluation, compare with Figure~\ref{field_comparison_measured_data}.

\section{Summary and conclusions}

We have investigated the possibility of using the ratio of the densities of the 2p states for reduced electric field strength ($E/N$) determination from the optical emission spectra for %atmospheric-pressure 
%argon streamer discharges. 
nanosecond pulsed barrier discharges in argon at atmospheric pressure. 
To find robust pairs of 2p states, 
we have used the 
%novel methodology by 
combined methodology of  
performing a large number of model calculations in 1D and 2D geometry 
%of the investigated nanosecond pulsed barrier discharge 
for different frequencies from 5\,kHz to 100\,kHz and evaluating the data at different spatial coordinates in the gap during streamer development using a semi-automated procedure. 
Such an approach enabled us to study the applicability of selected 2p states for the $E/N$ determination, based on a comparison of the calculated values with the original $E/N$ given by the 1D and 2D modelling results, respectively. 
The computations were done 
% for frequencies up to 100\,kHz 
for different positions in the streamer development to assess the sensitivity of the investigated 2p state ratios for $E/N$ to initial metastable densities (up to approx.\ 7$\times$10$^{15}$\,m$^{-3}$ at maximum),
electron densities (up to approx.\ 5$\times$10$^{17}$\,m$^{-3}$ at maximum) and for a wide range of $E/N$ values (70 to 2200\,Td). 
%\ja{\st{It should be emphasized that the main goal is to explore the sensitivity of the selected reaction kinetic processes regarding changes of the electric field and to find out which measurable 2p levels of argon are most suitable to re-calculate the electric field from measured optical emission data.}}

\ja{The presented work is a required and important step towards introducing the line intensity ratio method for experimental determination of the electric field in highly transient argon plasmas. This step consists of exploring the possibilities of the described methodology by means of numerical simulations and semi-automated evaluation of large amounts of data. The main purpose is to determine the sensitivity of selected reaction kinetic processes regarding changes of the electric field and to find out which measurable 2p levels of argon (in Paschen notation) are suitable to re-calculate the electric field from measured emission profiles. This can obviously be done only on the basis of numerical simulations providing the spatiotemporal development of all relevant 2p states, 
\ja{reaction rate coefficients}
%reaction rates 
as well as the electric field.}

Special attention was paid to identify appropriate effective lifetimes of selected 2p states, as these are an important part of 
%the equation for 
the $E/N$ determination. Two effective lifetime definitions were proposed, which are considered to be useful for the evaluation of measurements in the laboratory. 

Few ratios of 2p states were identified as the most robust ones. 
The density ratio 2p$_4$/2p$_9$ shows good results for both evaluated effective lifetimes and most of the positions and frequencies.
The computed $E/N$ values evaluated from the density ratio 2p$_2$/2p$_3$ do not show the best agreement with the simulated $E/N$, yet its ratio of excitation rate coefficients manifests a relatively good sensitivity to $E/N$.  
The most promising 2p states' ratio is 2p$_2$/2p$_5$. It even shows qualitative agreement if applied to measured data, even though its rate coefficients' ratio is not very sensitive to the $E/N$ value. \ja{This is a very promising result, clearly surpassing the expectations and overcoming the challenging limitation of the intensity ratio methods in general as they are currently known \cite{goldberg2022}. 
}

The results of the semi-automated procedure, in the form of the established $E/N$ multiplication coefficients, have shown that for  frequencies higher than 20\,kHz and the area at close vicinity of the 
\ja{dielectric}
%dielectrics 
on the cathode, the uncertainty of almost all 2p states' ratios rises and makes its use for $E/N$ determination far more complicated if not completely uncertain. \ja{This is mostly due to the increasing influence of indirect gain and loss processes for the 2p states, such as stepwise excitation.
 }

%Nevertheless, few 2p states' ratios were identified as the most robust ones. 
%The computed $E/N$ evaluated from the density ratio of 2p$_2$/2p$_3$ does not show the best agreement with the simulated $E/N$, yet its rate coefficients' ratio manifests a relatively good sensitivity to $E/N$. The density ratio of 2p$_4$/2p$_9$ shows good results for both evaluated effective lifetimes and most of the positions and frequencies. 
%However, the most promising 2p states' ratio showing also a qualitative agreement if applied to the experimental data is 2p$_2$/2p$_5$, even if its rate coefficients' ratio is not sensitive much to the $E/N$.

%Finally, we 
Furthermore, we investigated the effect of axial and radial integration of the simulated particle densities on the $E/N$ values calculated by the \ja{line intensity} ratio method. 
The uncertainty generally remains 
in a reasonable range below 20\%.  
It was shown as well 
that the axial integration of the signal (if the method is used in the experiment) over a few tens of microns is not as crucial as the radial integration. 
%Yet it remains still in a reasonable range under 20\% of uncertainty, given the evaluated ultra-fast phenomena.

The performed studies and analyses   
show the potential of 
quantifying the 2p states' ratios 
 for the direct use in plasma diagnostics 
 when combined with modelling studies. 
 They also serve for the validation of the fluid-Poisson modelling and the recently developed reaction kinetics scheme on sub-nanosecond and microscopic scales by direct comparison with experimental data of comparable spatiotemporal resolution. \ja{Such spatiotemporal scales of combined theoretical and direct experimental investigations were up to now mostly inaccessible. Moreover, the semi-automated procedure can be understood as a first step towards fully automated evaluation of large simulation data not only for plasma diagnostics.}

%Apparently, the obtained results do not give a simple solution and we do not offer a new diagnostics method for streamer discharges in argon. 
%Yet quantifying the 2p states' ratios and investigating all the expected challenging issues, we see such a method as possible, if the input data will be measured with high sensitivity and spatiotemporal resolution. 
%The performed analysis and theoretical effort also serve for validation of the fluid-Poisson model and the developed reaction kinetics on sub-nanosecond and microscopic scales by direct comparison with experimental data of comparable spatiotemporal resolution. We take it also as a first step towards a more automatized evaluation of theoretical data for direct use in plasma diagnostics. 

\section*{Acknowledgments}
This research was funded by the Czech Science Foundation project no.\,21-16391S and by the Deutsche Forschungsgemeinschaft (DFG, German Research Foundation) Projects No. 407462159 and 466331904. 
This research has been also supported by the project LM2023039 funded by the Ministry of Education, Youth and Sports of the Czech Republic.

\section*{References}
\bibliography{literature}
\bibliographystyle{iopart-num}

\end{document}